\documentstyle[12pt,epsf]{article}

\advance\voffset by -1.5cm
\advance\hoffset by -1.8cm
\def\be{\begin{equation}}
\def\ee{\end{equation}}

\def\Zop{\bbbz}
\def\Rop{\bbbr}

\def\N{{\cal N}}
\def\I{{\cal I}}
\def\half {\frac{1}{2}}
\def\bbbz {{\sf Z\!\!Z}}

\def\bbbr {{\rm I\!R}}

\def\ie {{\it i.e.}}
\newcommand{\ket}[1]{|#1\rangle}
\newcommand{\bra}[1]{\langle#1|}

\begin{document}
\thispagestyle{empty}
\def\thefootnote{\fnsymbol{footnote}}
\begin{flushright}
hep-th/0005029 \\
DAMTP-2000-9 
\end{flushright}
\vspace{2.0cm}

\begin{center}

{\Large {\bf Lectures on Non-BPS Dirichlet branes}}\footnote{Lectures
given at the TMR network school on `Quantum aspects of gauge theories,
supersymmetry and quantum gravity', Torino, 26 January  -- 2 February
2000, and at the `Spring workshop on Superstrings and related matters',
Trieste, 27 March -- 4 April 2000.}
\vspace{2.0cm}

{\large Matthias R. Gaberdiel} 
\footnote{{\tt E-mail: M.R.Gaberdiel@damtp.cam.ac.uk}} \\
\vspace{0.2cm} 

{Department of Applied Mathematics and Theoretical Physics\\ 
University of Cambridge,  \\
Wilberforce Road, \\
Cambridge CB3 0WA, U.\ K.\ }
\vspace{0.5cm}

\vspace{0.5cm}

May 2000
\vspace{3.0cm}

{\bf Abstract}
\end{center}

{\leftskip=2.4truecm
\rightskip=2.4truecm

A comprehensive introduction to the boundary state approach to
Dirichlet branes is given. Various examples of BPS and non-BPS
Dirichlet branes are discussed. In particular, the non-BPS states in
the duality of Type IIA on K3 and the heterotic string on $T^4$ are
analysed in detail.

}

\newpage

\setcounter{footnote}{0}
\def\thefootnote{\arabic{footnote}}

\section{Introduction}
\renewcommand{\theequation}{1.\arabic{equation}}
\setcounter{equation}{0}

The past few years have seen a tremendous increase in our
understanding of the dynamics of superstring theory. In particular it
has become apparent that the five ten-dimensional theories, together
with an eleven-dimensional theory (M-theory), are different limits in
moduli space of some unifying description.  A crucial ingredient in
understanding the relation between the different perturbative
descriptions has been the realisation that the {\em solitonic} objects
that define the relevant degrees of freedom at strong coupling are
{\em Dirichlet-branes} that have an alternative description in terms
of open string theory \cite{Dai,mbg1,Pol1}. 

The D-branes that were first analysed were BPS states that break half
the (spacetime) supersymmetry. It has now been realised, however, that
because of their description in terms of open strings, D-branes can be
constructed and analysed in much more general situations. In fact,
D-branes are essentially described by a boundary conformal field
theory \cite{PolCai,CLNY,Li,GrGut,Cardy,Lew,PSS1,PSS2} (see also
\cite{Love71,ClaSha,AAGNSV,CLNY0,BiaSag,Hor} for earlier work in this
direction), the consistency  conditions of which are not related to
spacetime supersymmetry \cite{BG1,BG2,KT} (for an earlier
non-supersymmetric orientifold construction see also
\cite{BiaSag}). In an independent development, D-branes that break
supersymmetry have been constructed in terms of bound states of branes
and anti-branes by Sen \cite{Sen1,Sen2,Sen3,Sen4,Sen5,Sen6} (see also
\cite{SenRev} for a good review). This beautiful construction has been
interpreted in terms of K-theory by Witten \cite{WittenK}, and this
has opened the way for a more mathematical treatment of D-branes 
\cite{Horava,Gukov,BGH}. It has also led the way to new insights into
the nature of the instability that is described by the open string
tachyon \cite{stringfield}.

The motivation for studying D-branes that do not preserve spacetime
supersymmetry (and that are therefore sometimes called non-BPS
D-branes) is at least four-fold.  First, in order to understand the
strong/weak coupling dualities of supersymmetric string theories in
more detail, it is important to analyse how these dualities act on
states that are not BPS saturated. After all, the behaviour of the BPS
states at arbitrary coupling is essentially determined by spacetime
supersymmetry (provided that it remains unbroken for all values of the
coupling constant), and thus one is not really probing the underlying
string theory unless one also understands how non-BPS states behave at
strong coupling. The dualities typically map perturbative states to
non-perturbative (D-brane type) states, and thus one will naturally
encounter non-BPS D-branes in these considerations.

The second motivation is related to the question of whether string
duality should intrinsically only apply to supersymmetric string
theories, or whether also non-super\-sym\-met\-ric theories should be
related by duality. This is certainly, {\it a priori}, an open
question\footnote{Recently, some suggestive proposals have however
been made \cite{BG1,BD,BG4,BK,KKS,KS,Harvey}.}: it is conceivable that
spacetime supersymmetry is a crucial ingredient without which there is
no reason to believe that these dualities should exist, but it is also
conceivable that spacetime supersymmetry is just a convenient tool
that allows one to use sophisticated arguments and techniques to
verify conjectures that are otherwise difficult to check. Dirichlet
branes play a central r\^ole in the understanding of string dualities,
and if one wants to make progress on this question, it is important
to develop techniques to analyse and describe Dirichlet branes without 
reference to spacetime supersymmetry. 

Thirdly, one of the interesting implications of the Maldacena
conjecture \cite{Maldacena} is that one can obtain non-trivial
predictions about field theory from string theory. In the original
formulation this was applied to supersymmetric string and field
theories, but it is very tempting to believe that similar insights may
be gained for non-supersymmetric theories. This line of thought has
been developed recently, starting with a series of papers by 
Klebanov \& Tseytlin \cite{KT}. 

Finally, non-BPS D-branes offer the intriguing possibility of string
compactifications in which supersymmetry is preserved in the bulk but
broken on the brane. Various orientifold models involving branes and
anti-branes have been constructed recently \cite{ADS,AU,AIQ,AADDS}, 
but it is presumably also possible to construct interesting
models involving non-BPS D-branes. (Non-BPS D-branes in Type II
theories have recently been considered in \cite{Mukhi1,Mukhi2}.) The
fact that at specific points in the moduli space their spectrum is
Bose-Fermi degenerate may be of significance in this context
\cite{GabSen}.   
\bigskip

The main aim of these lectures is to explain the boundary state
approach to D-branes, and to give some applications of it, in
particular to the construction of non-BPS D-branes. The structure of
the lectures is as follows. In section~2 we explain carefully the
underpinnings of the boundary state approach and apply it to the
simplest case, Type IIA/IIB and Type 0A/0B. In section~3 we use the
techniques that we have developed to construct one of the simplest
non-BPS D-branes --- the non-BPS D-particle of the orbifold of Type
IIB by $(-1)^{F_L} \I_4$ --- in detail. If we compactify this theory
on a 4-torus, it is T-dual to Type IIA at the orbifold point of K3,
which in turn is S-dual to the heterotic string on $T^4$. This
connection (and in particular the various non-BPS states in this
duality) are analysed in detail in section~4.

\section{The boundary state approach}
\renewcommand{\theequation}{2.\arabic{equation}}
\setcounter{equation}{0}

Suppose we are given a closed string theory. We can ask the question
whether it is possible to add to this theory additional open string
sectors in such a way that the resulting open- and closed theory is
consistent. The different open string sectors that we can add are
characterised by the {\em boundary conditions} that we impose on the
end-points of the open strings. Conventional open strings have 
{\em Neumann} boundary conditions at either end; if we denote by
$X^\mu(t,s)$ the coordinate field, where $t\in\Rop$ and $s\in[0,\pi]$
are the time and space coordinates on the world-sheet of the open
string, then this is the condition that 
\be\label{Neumann}
\partial_s X^\mu(t,0) = 
\partial_s X^\mu(t,\pi) = 0 \,.
\ee
We can also consider open strings whose boundary condition at one or
both ends is of {\em Dirichlet} type, \ie\ 
\be\label{Dirichlet}
X^\nu(t,0) = a^\nu \,,
\ee
where $a^\nu$ is a constant. Finally, we can consider open strings
that satisfy Neumann boundary conditions for some of the $X^\mu$,  
and Dirichlet boundary conditions for the others 
\be\label{openc}
\begin{array}{rcll}
\partial_s X^\mu(t,0) & = & 0 \qquad &
\hbox{$\mu=0,\ldots, p$} \\
X^\nu(t,0) & = &  a^\nu \quad &
\hbox{$\nu=p+1,\ldots,9$}\,.
\end{array}
\ee
The endpoint of such an open string is then constrained to lie on a
submanifold (a hyperplane of dimension $p+1$), whose position in the
ambient space is described by $a^\nu$; this submanifold is then called
the Dirichlet $p$-brane or $Dp$-brane for short. The different
boundary conditions of the open string are in one-to-one
correspondence with the different D-branes.  We can therefore rephrase
the above question as the question of which D-branes can be
consistently defined in a given closed string theory.   

The idea of the boundary state approach to D-branes is to represent a
D-brane as a coherent (boundary) state of the underlying closed string
theory. The key ingredient in this approach is {\em world-sheet
duality} that allows one to rewrite the above conditions (that are
defined in terms of the coordinate function of the open string) in
terms of the coordinate function of the closed string. At first, the
coordinate functions of the open and the closed string theory are not
related at all: the world-sheet of the open string is an infinite
strip, whilst the world-sheet of the closed string has the topology of
a cylinder. For definiteness, let us parametrise the closed string
world-sheet by $\tau$ and $\sigma$, where $\tau\in\Rop$ is the 
time variable, and $\sigma$ is a {\em periodic} space-variable
$\sigma\in [0,2\pi]$ (where $\sigma=0$ is identified with
$\sigma=2\pi$).  

Suppose now that we consider an open string that has definite boundary
conditions at either end (and can therefore be thought of as
stretching between two not necessarily different D-branes). If we
determine the 1-loop partition function of this open string, we have
to identify the time coordinate periodically (and integrate over all
periodicities). The open-string world-sheet has then the topology of a
cylinder, where the periodic variable is $t$, and $s$ takes values in a
finite interval (from $s=0$ to $s=\pi$). Because of world-sheet
duality, we can re-interpret this world-sheet as being a closed string
world-sheet if we identify $t$ with $\sigma$ (up to normalisation) and
$s$ with $\tau$. From the point of view of the closed string, the
diagram then corresponds to a tree-diagram between two external
states; this describes the processes, where closed string states are
emitted by one external state and absorbed by the other.
\begin{figure}[h]
\hspace*{2.5cm}
\epsfxsize=10cm
\epsfbox{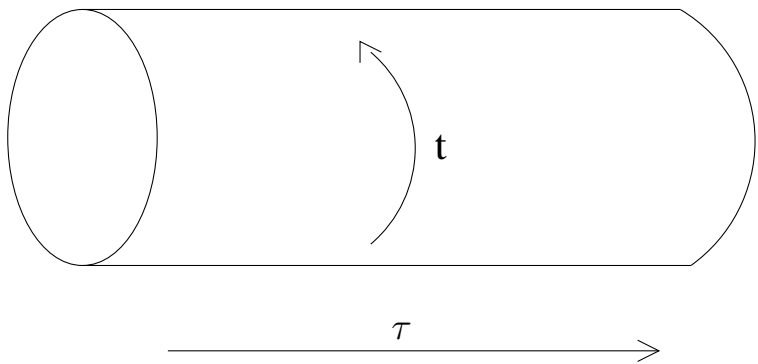}
\caption{World-sheet duality}
\end{figure}

\vspace*{-2.2cm}
\hspace{7cm}
{\large $\tau$}
\vspace{2cm}

The boundary condition on the ends of the open string become now 
conditions that must be satisfied by the external states; since we
exchange $(t,s)$ with $(\sigma,\tau)$ these conditions are then
\be\label{closedc}
\begin{array}{rcll}
\partial_\tau X^\mu(\sigma,0) |Dp\rangle& = & 0 \qquad &
\hbox{$\mu=0,\ldots, p$} \\
X^\nu(\sigma,0) |Dp\rangle & = &  a^\nu |Dp\rangle \quad &
\hbox{$\nu=p+1,\ldots,9$}\,,
\end{array}
\ee
and a similar relation for $s=\tau=\pi$. Here we have assumed that the 
boundary condition at $s=\tau=0$ corresponds to those of a $Dp$-brane.  

It is useful to rewrite these conditions in terms of the modes of the
closed string theory. To this end, let us recall that the coordinate
field in the closed string theory can be written as 
\be
X^\mu(\tau,\sigma) = X^\mu_L(\tau+\sigma) + X^\mu_R(\tau-\sigma)\,,
\ee
where in terms of modes,
\begin{eqnarray}
X^\mu_L & = & \half x^\mu + \half p^\mu (\tau+\sigma) + 
{i \over 2} \sum_{n\ne 0} {1\over n} \alpha^\mu_n
e^{-in(\tau+\sigma)} \\
X^\mu_R & = & \half x^\mu + \half p^\mu (\tau-\sigma) + 
{i \over 2} \sum_{n\ne 0} {1\over n} \widetilde\alpha^\mu_n
e^{-in(\tau-\sigma)}\,.
\end{eqnarray}
These modes satisfy the commutation relations
\be
\begin{array}{rcl}
{}[\alpha^\mu_m,\alpha^\nu_n] & = & m \,\delta^{\mu\nu}\, \delta_{m,-n}
\\
{}[\alpha^\mu_m,\widetilde\alpha^\nu_n] & = & 0 \\
{}[\widetilde{\alpha}^\mu_m,\widetilde{\alpha}^\nu_n] & = & 
m \, \delta^{\mu\nu} \, \delta_{m,-n} \,.
\end{array}
\ee
In terms of modes the conditions (\ref{closedc}) then become
\be\label{ff1}
\begin{array}{rcll}
p^\mu |Dp\rangle & 
= & 0  \qquad & \hbox{$\mu=0,\ldots, p$} \\
\left(\alpha^\mu_n + \widetilde\alpha^\mu_{-n} \right) |Dp\rangle & 
= & 0  \qquad & \hbox{$\mu=0,\ldots, p$} \\
\left(\alpha^\nu_n - \widetilde\alpha^\nu_{-n} \right) |Dp\rangle & 
= & 0 \qquad & \hbox{$\nu=p+1,\ldots,9$} \\
x^\nu |Dp\rangle & = & a^\nu |Dp\rangle \quad & 
\hbox{$\nu=p+1,\ldots,9$}\,.
\end{array}
\ee

The boundary conditions for the fermions are more difficult to
establish. Ultimately they are determined by the condition that the
closed string tree diagram reproduces, upon world-sheet duality, the
open string loop diagram (see (\ref{++}) -- (\ref{--})
below). In order to formulate the appropriate condition, it is
necessary to introduce an additional parameter $\eta$ (that
corresponds to the different spin structures), and the relevant
equations are then
\be\label{fermionmodes}
\begin{array}{rcll}
\left(\psi^\mu_r + i \eta \widetilde\psi^\mu_{-r} \right) 
|Bp,\eta\rangle & 
= & 0  \qquad & \hbox{$\mu=0,\ldots, p$} \\
\left(\psi^\nu_r - i \eta \widetilde\psi^\nu_{-r} \right) 
|Bp,\eta\rangle & 
= & 0 \qquad & \hbox{$\nu=p+1,\ldots,9$} \,.
\end{array}
\ee
The actual D-brane state $|Dp\rangle$ is then a linear combination of
the boundary states $|Bp,\eta\rangle$ with $\eta=\pm$. It is also
worth pointing out that the equations can be solved separately for the
different closed string sectors of the theory (\ie\ the NS-NS and the
R-R sector, as well as the corresponding twisted sectors if we are dealing
with an orbifold theory). We shall therefore, in the following,
usually denote by $|Bp,\eta\rangle$ the solution in a specific sector
of the theory; the D-brane state is then a certain linear combination
of the boundary states in the different sectors and with $\eta=\pm$. 

In the following we shall always work in the NS-R formalism; we shall
also, for simplicity, work in light-cone gauge, and we shall always
choose the two light-cone directions to be $\mu=0,1$.\footnote{For a
good introduction to the covariant approach see the lecture notes by 
Di Vecchia and Liccardo \cite{VecLic}.} The boundary
conditions in both light-cone directions will be taken to be
Dirichlet; the boundary states we describe are therefore really
D-instantons (\ie\ they satisfy a Dirichlet boundary condition in 
the time direction). However, by means of a double Wick rotation,
these states can be transformed into states whose boundary conditions
are specified as above \cite{GrGut}. In these conventions 
we necessarily restrict ourselves to D-branes with at least two
Dirichlet directions; thus we can only describe $Dp$-branes with 
$-1\leq p\leq 7$. Also, since the two light-cone directions are always
Dirichlet, only $p-2$ of the transverse directions satisfy a Dirichlet
boundary condition in order for the state to describe the Wick rotate
of a $Dp$-brane; thus in these conventions the boundary states that
combine to define a $Dp$-brane are characterised by the following
conditions 
\be\label{Dplc}
\begin{array}{rcll}
p^\mu |Bp,\eta\rangle & 
= & 0  \qquad & \hbox{$\mu=2,\ldots, p+2$} \\
\left(\alpha^\mu_n + \widetilde\alpha^\mu_{-n} \right) 
|Bp,\eta\rangle & 
= & 0  \qquad & \hbox{$\mu=2,\ldots, p+2$} \\
\left(\alpha^\nu_n - \widetilde\alpha^\nu_{-n} \right) 
|Bp,\eta\rangle & 
= & 0 \qquad & \hbox{$\nu=p+3,\ldots,9$} \\
x^\nu |Bp,\eta\rangle & = & a^\nu |Bp,\eta\rangle \quad & 
\hbox{$\nu=0,1,p+3,\ldots,9$} \\
\left(\psi^\mu_r + i \eta \widetilde\psi^\mu_{-r} \right) 
|Bp,\eta\rangle & 
= & 0  \qquad & \hbox{$\mu=2,\ldots, p+2$} \\
\left(\psi^\nu_r - i \eta \widetilde\psi^\nu_{-r} \right) 
|Bp,\eta\rangle & 
= & 0 \qquad & \hbox{$\nu=p+3,\ldots,9$} \,.
\end{array}
\ee

It is actually not difficult to describe the solution to these
equations. In each (left-right-symmetric) sector of the theory, and
for each choice of $\eta$, the unique solution is of the form
\be\label{114}
|Bp,{\bf a},\eta\rangle = {\cal N} \int 
\prod_{\nu=0,1,p+3,\ldots, 9} dk^\nu e^{i k^\nu a^\nu} 
\widehat{|Bp,{\bf k},\eta\rangle}\,,
\ee
where ${\cal N}$ is a normalisation constant that will be determined
further below, and $\widehat{|Bp,{\bf k},\eta\rangle}$ is the coherent
state 
\begin{eqnarray}\label{115}
\widehat{|Bp,{\bf k},\eta\rangle} & = & \exp\left\{ \sum_{n>0} 
\left[ -{1\over n} \sum_{\mu=2}^{p+2} 
	\alpha^\mu_{-n} \widetilde\alpha^\mu_{-n} 
+ {1\over n} \sum_{\nu=p+3}^{9} 
	\alpha^\nu_{-n} \widetilde\alpha^\nu_{-n} \right] \right.\\
& & \qquad \qquad \left. + i \eta \sum_{r>0}
\left[ - \sum_{\mu=2}^{p+2} 
	\psi^\mu_{-r} \widetilde\psi^\mu_{-r} 
+ \sum_{\nu=p+3}^{9} \psi^\nu_{-r} \widetilde\psi^\nu_{-r} 
\right] \right\} |Bp,{\bf k},\eta\rangle^{(0)}\nonumber \,.
\end{eqnarray}
The ground state is a momentum eigenstate with eigenvalue 
${\bf k}$, where $k^\mu=0$ for $\mu=2,\ldots, p+2$; in the NS-NS
sector, it is the unique tachyonic ground state, whereas in the R-R
sector, it is determined by the condition  (\ref{fermionmodes}) with
$r=0$, \ie\ 
\be\label{fermionzero}
\begin{array}{lcll}
\left(\psi^\mu_0 + i \eta \widetilde\psi^\mu_{0} \right) 
|Bp,{\bf k},\eta\rangle^{(0)}_{\mbox{\tiny{R-R}}} & = & 0  \qquad & 
\hbox{$\mu=2,\ldots, p+2$} \\
\left(\psi^\nu_0 - i \eta \widetilde\psi^\nu_{0} \right) 
|Bp,{\bf k},\eta\rangle^{(0)}_{\mbox{\tiny{R-R}}} & = & 0  \qquad & 
\hbox{$\nu=p+3,\ldots,9$} \,.
\end{array}
\ee
If the theory under consideration is an orbifold theory (such as the
theory we shall discuss later), there are also similar boundary states
in the corresponding twisted sectors. The actual D-brane state is then
a certain linear combination of these states in the different sectors
of the theory and for both values of $\eta$; it is characterised by
three properties \cite{BG1,BG4}:
\begin{list}{(\roman{enumi})}{\usecounter{enumi}}
\item The boundary state only couples to the physical sector of the
closed string theory, {\it i.e.} it is GSO-invariant, and invariant
under orbifold and orientifold projections where appro\-pri\-ate.
\item The open string amplitude obtained by world-sheet duality
from the closed string exchange between any two boundary states 
constitutes an {\em open string partition function}, {\it i.e.} 
it corresponds to a trace over a set of open string states
of the open string time-evolution operator.
\item The open strings that are introduced in this way have 
{\em consistent string field interactions} with the original closed
strings. 
\end{list}
One is usually also interested in D-branes that are {\em stable}; 
a necessary condition for this is that the spectrum of open strings
that begin and end on the {\em same} D-brane is free of tachyons. If
the underlying theory is supersymmetric, one may sometimes also want
to impose the condition that the D-branes preserve some part of the
supersymmetry, and that they are therefore {\em BPS saturated}; this
requires that the spectrum of open strings beginning and ending on the
D-brane is supersymmetric. However, there exist interesting D-branes in
supersymmetric theories that are stable but not BPS
\cite{Sen2,Sen3,BG2,WittenK,BG3,FGLS,SenRev,LR,GabSte,DS}; some
examples of these will be described later.  

The first condition is usually relatively easy to check, although it
requires care in all sectors that have fermionic zero modes. (We shall
describe the relevant subtleties in some detail for the case of Type
IIA and Type IIB in subsection 2.2.) The second condition is in
essence equivalent to the statement that world-sheet duality holds.
It is a very powerful constraint that determines the normalisations of 
the different boundary states (as we shall show in the next
subsection). This condition can be formulated in terms of the
conformal field theory and is sometimes referred to as Cardy's
condition (see also subsection 2.4). The third condition is very
difficult to check in detail; as far as I am aware, there is only one 
example (namely the two Type 0 theories) for which it seems to
imply constraints that go beyond (i) and (ii). 

The set of boundary states which satisfies these conditions forms a 
{\em lattice}. This follows from the fact that if the set of boundary
states ${\cal S}=\{|D\rangle_1, |D\rangle_2, \ldots\}$ satisfies these
conditions, then so will the set of boundary states that contains in
addition to the elements of ${\cal S}$ any integer-valued linear
combination of $|D\rangle_1, |D\rangle_2,\ldots$. When we talk about 
the D-branes of a theory, what we really mean are the {\em basis
vectors} of this lattice, from which every D-brane of the theory can
be obtained as an integer-valued linear combination; this is what we
shall determine in the following.

In general, a given theory can have different lattices of mutually
consistent boundary states that are not consistent relative to each
other. In this case, condition (iii) presumably selects the correct
lattice of boundary states. (This is at least what happens in the case
of the Type 0 theories.) 

Finally, it should be stressed that the above conditions are 
{\em intrinsic consistency conditions} of an interacting string
(field) theory; in particular, they are more fundamental than
spacetime supersymmetry, and also apply in cases where spacetime
supersymmetry is broken or absent.

\subsection{World-sheet duality}

Before describing some examples in detail, it is useful to illustrate
the condition of world-sheet duality more quantitatively (since the
same calculation will be needed for essentially all models). The
closed string tree diagram that is represented in Figure~1 is
described by  
\be\label{closedtree}
\int_0^\infty dl\, \langle Dq | e^{-l H_c} \, | Dp \rangle \,,
\ee
where $H_c$ is the closed string Hamiltonian in light cone gauge,
\be
H_c = \pi {\bf k}^2 + 2\pi \sum_{\mu=2,\ldots,9} \left[
\sum_{n=1}^{\infty} (\alpha^\mu_{-n} \alpha^\mu_{n}
      + \tilde{\alpha}^\mu_{-n} \tilde{\alpha}^\mu_{n})
+ \sum_{r>0} r (\psi^\mu_{-r} \psi^\mu_r
      + \tilde\psi^\mu_{-r}\tilde\psi^\mu_{r} ) \right] + 2 \pi C_c
\,.
\ee
The constant $C_c$ takes the value $-1$ in the NS-NS, and $0$ in the
R-R sector. Under the substitution $t=1/2l$, this integral should become 
the open string one-loop amplitude that is given by 
\be
\int_0^\infty {dt \over 2t} \hbox{Tr} (e^{-2 t H_o} {\cal P}) \,,
\ee
where ${\cal P}$ is an appropriate projection operator, and $H_o$ is
the open string Hamiltonian
given as 
\be \label{ex2}
H_o= \pi \vec{p}^2 + {1 \over 4\pi} \vec{w}^2 
+ \pi \sum_{\mu =2,\ldots, 9 }  
[\sum_{n=1}^\infty \alpha^\mu_{-n} \alpha^\mu_n + \sum_{r>0} r
\psi^\mu_{-r} \psi^\mu_r] + \pi C_o\, .
\ee
Here $\vec{p}$ denotes the open string momentum along the directions for
which the string has Neumann (N) boundary conditions at both ends,
$\vec{w}$ is the difference between the two end-points of the open
string, and $\alpha^\mu_n$ and $\psi^\mu_r$ are the bosonic and
fermionic open string oscillators, respectively; they satisfy the
commutation relations 
\be \label{ex3}
[\alpha^\mu_m,\alpha^\nu_n]= m\,\delta^{\mu\nu}\,\delta_{m,-n}, \qquad
\{\psi^\mu_r, \psi^\nu_s\}=\delta^{\mu\nu}\,\delta_{r,-s}\, .
\ee
For coordinates satisfying the same boundary condition at both ends of
the open string ({\it i.e.} both Neumann (N) or both Dirichlet (D)) 
$n$ always takes integer values, whereas $r$ takes integer (integer + 
${1\over 2}$) values in the R (NS) sector. On the other hand, for
coordinates satisfying different boundary conditions at the two ends
of the open string (one D and one N) $n$ takes integer+${1\over 2}$
values and $r$ takes integer +${1\over 2}$ (integer) values in the R
(NS) sector. The normal ordering constant $C_o$ vanishes in the
R-sector and is equal to $-{1\over 2}+{s\over 8}$ in the NS sector (in
$\alpha'=1$ units) where $s$ denotes the number of coordinates
satisfying D-N boundary conditions. The trace, denoted by $\hbox{Tr}$,
is taken over the full Fock space of the open string, and also
includes an integral over the various momenta. 

The calculation (\ref{closedtree}) can be performed separately for the
different boundary states in the different components since the
overlap between states from different sectors vanishes. For
definiteness let us consider one specific example in some detail, the
tree exchange between two $Dp$-brane boundary states in the NS-NS
sector. (The result for the other sectors will be given below.) Thus 
we want to consider the amplitude 
\be
\int_0^\infty dl\, \langle Bp,{\bf a_1},\eta | e^{-l H_c} \, 
| Bp,{\bf a}_2, \eta \rangle_{\mbox{\tiny{NS-NS}}} \,,
\ee
where $| Bp,{\bf a}, \eta \rangle_{\mbox{\tiny{NS-NS}}}$ is the
coherent state in the NS-NS sector given in (\ref{114}). The momentum
integral gives a Gaussian integral that can be performed, and the
amplitude becomes\footnote{The amplitude is bilinear in the external
states, and the prefactor is therefore ${\cal N}^2$ rather than 
${\cal N}\overline{\cal N}$. On momentum eigenstates the amplitude
satisfies $\langle k_1 | k_2 \rangle=\delta(k_1+k_2)$.} 
\be
{\cal N}_{\mbox{\tiny{NS-NS}}}^2 \int_0^\infty dl \; 
l^{-{9-p \over 2}} e^{-{({\bf a}_1 - {\bf a}_2)^2 \over 4\pi l}} \;
\langle \widehat{Bp,{\bf 0},\eta} | e^{-l H_c} | 
\widehat{Bp,{\bf 0},\eta}\rangle_{\mbox{\tiny{NS-NS}}}\,.
\ee
In order to determine the overlap between the two coherent states, we
observe that the states of the form
\be
\prod_i {1 \over l_i!} \left({1\over n_i} \alpha^{\mu_i}_{-n_i}
\widetilde{\alpha}^{\mu_i}_{-n_i} \right)^{l_i} |0\rangle \,,
\ee
where $n_i\geq n_{i+1}$ and if $n_i=n_{i+1}$ then $\mu_i<\mu_{i+1}$,
form an {\em orthonormal basis} for the space generated by the modes
$\alpha^\mu_n \widetilde{\alpha}^\mu_n$ and similarly for 
\be
\prod_i {1\over l_i!} \left(i \eta \psi^{\mu_i}_{-r_i}
\widetilde{\psi}^{\mu_i}_{-r_i} \right)^{l_i} |0\rangle \,.
\ee
Here we have used that the bilinear inner product is defined by the
relation 
\be
\langle \alpha^\mu_n \phi | \chi \rangle
= - \langle\phi| \alpha^\mu_{-n}\chi\rangle\,, \qquad
\langle\psi^\mu_n \phi|\chi\rangle
= i \langle\phi| \psi^\mu_{-n}\chi\rangle\,, 
\ee
and similarly for $\widetilde{\alpha}^\mu_n$ and
$\widetilde{\psi}^\mu_r$ together with the normalisation
\be
\Bigl\langle |0\rangle \Bigl|  |0 \rangle \Bigr\rangle = 1 \,.
\ee
It is then easy to see that the above amplitude becomes
\be\label{form1}
{\cal N}_{\mbox{\tiny{NS-NS}}}^2 \int_0^\infty dl \; l^{-{9-p \over 2}} 
e^{-{({\bf a}_1 - {\bf a}_2)^2 \over 4\pi l}} \; 
{f_3^8(q) \over f_1^8(q)}  \,,
\ee
where $q=e^{-2\pi l}$, and the functions $f_i$ are defined as in  
\cite{PolCai} 
\begin{eqnarray} \label{ffn}
f_1(q) & = & q^{1\over 12} \prod_{n=1}^\infty
( 1 - q^{2n})\,, \nonumber
\\
f_2(q) & = & \sqrt 2 q^{1\over 12}
\prod_{n=1}^\infty ( 1 + q^{2n})\,,
\nonumber
\\
f_3(q) & = & q^{-{1\over 24}}
\prod_{n=1}^\infty ( 1 + q^{2n-1})\,,
\nonumber
\\
f_4(q) & = & q^{-{1\over 24}}
\prod_{n=1}^\infty ( 1 - q^{2n-1})\,.
\end{eqnarray}

Next we substitute $t=1/2l$, and using the transformation properties
of the $f_i$ functions,
\be\label{trans}
\begin{array}{rclrcl}
f_1(e^{-\pi/t}) & = & \sqrt{t} f_1(e^{-\pi t})\,, \qquad &
f_2(e^{-\pi/t}) & = & f_4(e^{-\pi t})\,, \\
f_3(e^{-\pi/t}) & = & f_3(e^{-\pi t})\,, \qquad &
f_4(e^{-\pi/t}) & = & f_2(e^{-\pi t})\,,
\end{array}
\ee
the above integral becomes
\be
{\cal N}_{\mbox{\tiny{NS-NS}}}^2 2^{9-p \over 2} \int_0^\infty 
{dt \over 2t} t^{-{(p+1) \over 2}} 
e^{-{({\bf a}_1 - {\bf a}_2)^2 \over 2\pi }t} \; 
{f_3^8(\tilde{q}) \over f_1^8(\tilde{q})}  \,,
\ee
where $\tilde{q}=e^{-\pi t}$. This is to be compared with the
open string one-loop amplitude
\be
\int_{0}^\infty {dt \over 2t} \hbox{Tr}_{\mbox{\tiny{NS}}} (e^{-2t H_o})
= {V_{p+1} \over (2\pi)^{p+1}} \int_{0}^\infty {dt \over 2t} 
(2t)^{-{(p+1) \over 2}} e^{-{({\bf a}_1 - {\bf a}_2)^2 \over 2\pi }t} \; 
{f_3^8(\tilde{q}) \over f_1^8(\tilde{q})} \,,
\ee
where $V_{p+1}$ is the world-volume of the brane, which together with
the factor of $(2t)^{-{(p+1) \over 2}}$ comes from the momentum
integration. Thus we find that 
\be\label{++}
  \int dl\, \bra{Bp,\eta}\, e^{-lH_{c}}\,
   \ket{Bp,\eta}_{\mbox{\tiny{NS-NS}}} 
 =  {\cal N}^2_{\mbox{\tiny{NS-NS}}} {32 (2\pi)^{p+1} \over V_{p+1}} 
      \int {dt\over 2t}\, \mbox{Tr}_{\mbox{\tiny{NS}}} 
      \Big[e^{-tH_{o}}\Big] \,.
\ee
Similarly we have
\be\label{+-}
  \int dl\, \bra{Bp,\eta}\, e^{-lH_{c}}\,
   \ket{Bp,-\eta}_{\mbox{\tiny{NS-NS}}} 
 =  {\cal N}^2_{\mbox{\tiny{NS-NS}}} {32 (2\pi)^{p+1} \over V_{p+1}} 
      \int {dt\over 2t}\, \mbox{Tr}_{\mbox{\tiny{R}}} 
       \Big[e^{-tH_{o}}\Big] \,,
\ee
\be\label{-+}
  \int dl\, \bra{Bp,\eta}\, e^{-lH_{c}}\,
   \ket{Bp,\eta}_{\mbox{\tiny{R-R}}} 
 =  - {{\cal N}^2_{\mbox{\tiny{R-R}}}\over 16}
      {32 (2\pi)^{p+1} \over V_{p+1}} 
      \int {dt\over 2t}\, \mbox{Tr}_{\mbox{\tiny{NS}}} 
      \Big[(-1)^F e^{-tH_{o}}\Big] \,,
\ee
and
\be\label{--}
  \int dl\, \bra{Bp,\eta}\, e^{-lH_{c}}\,
   \ket{Bp,-\eta}_{\mbox{\tiny{R-R}}} = 0 
=  - {{\cal N}^2_{\mbox{\tiny{R-R}}}\over 16}
      {32 (2\pi)^{p+1} \over V_{p+1}} 
      \int {dt\over 2t}\, \mbox{Tr}_{\mbox{\tiny{R}}} 
        \Big[(-1)^F e^{-tH_{o}}\Big] \,.
\ee

We learn from this that we can satisfy world-sheet duality provided we
include appropriate combinations of boundary states and choose their
normalisations correctly. We have now assembled the necessary
ingredients to work out some examples in detail.

\subsection{A first example: Type IIA and IIB}

Let us first consider the familiar case of the Type IIA and Type IIB
theories. The spectra of these theories is given by
\be
\begin{array}{ll}
 {\bf IIA}: & (\mbox{NS}+,\mbox{NS}+)\oplus (\mbox{R}+,\mbox{R}-)
           \oplus (\mbox{NS}+,\mbox{R}-)\oplus (\mbox{R}+,\mbox{NS}+) 
    \label{IIAGSO} \\[5pt]
 {\bf IIB}: & (\mbox{NS}+,\mbox{NS}+)\oplus (\mbox{R}+,\mbox{R}+)
           \oplus (\mbox{NS}+,\mbox{R}+)\oplus (\mbox{R}+,\mbox{NS}+)\,,
\end{array}
\label{typeIIspec}
\ee
where the signs refer to the eigenvalues of $(-1)^{F}$ and
$(-1)^{\widetilde{F}}$, respectively. In particular, the NS-NS sector is
the same for the two theories, and consists of those states for which
both $(-1)^{F}$ and $(-1)^{\widetilde{F}}$ have eigenvalue $+1$. Given 
that the tachyonic ground state has eigenvalue $-1$ under both
$(-1)^{F}$ and $(-1)^{\widetilde{F}}$, the boundary state given by
(\ref{114}) and (\ref{115}) transforms as 
\begin{eqnarray}
(-1)^{F} | Bp,{\bf a},\eta\rangle_{\mbox{\tiny{NS-NS}}}
 & = & - | Bp,{\bf a},-\eta\rangle_{\mbox{\tiny{NS-NS}}}
\nonumber \\
(-1)^{\tilde{F}} | Bp,{\bf a},\eta\rangle_{\mbox{\tiny{NS-NS}}}
 & = & - | Bp,{\bf a},-\eta\rangle_{\mbox{\tiny{NS-NS}}}
\nonumber\,.
\end{eqnarray}
Thus 
\be
\label{NSNS}
|Bp,{\bf a}\rangle_{\mbox{\tiny{NS-NS}}} 
= \left( |Bp,{\bf a},+\rangle{\mbox{\tiny{NS-NS}}} 
- |Bp,{\bf a},-\rangle{\mbox{\tiny{NS-NS}}} \right)
\ee
is a GSO-invariant state for all $p$. It follows from (\ref{++}) and
(\ref{+-}) that this state does not describe a {\em stable} D-brane by
itself since the open string that begins and ends on
$|Bp,{\bf a}\rangle_{\mbox{\tiny{NS-NS}}}$ consists of an unprojected
NS and R sector, and therefore contains a tachyon in its spectrum. In
fact (\ref{NSNS}) 
with 
\be\label{unstnorm}
{\cal N}^2_{\mbox{\tiny{NS-NS}}} (\widehat{Dp})= {1\over 64} 
{V_{p+1} \over (2\pi)^{p+1}} 
\ee
describes the unstable $\widehat{Dp}$-brane for $p$ odd (even) in Type
IIA (IIB) that was considered by Sen in his construction of non-BPS
D-branes \cite{Sen6,SenRev}; the unstable $\widehat{D9}$-brane of Type
IIA was also used by Ho\v{r}ava in his discussion of the K-theory of
Type IIA \cite{Horava}. 

In order to obtain a stable D-brane, we have to add to (\ref{NSNS}) a
boundary state in the R-R sector; since the R-R sector involves
fermionic zero modes, the discussion of GSO-invariance is somewhat
delicate, and we need to introduce a little bit of notation. Let us
define the modes
\be
\psi_{\pm}^{\mu} = {1 \over \sqrt{2}} 
\left(\psi_0^{\mu} \pm i \widetilde{\psi}_0^{\mu}\right)\,,
\ee
which satisfy the anti-commutation relations 
\be\label{fermanti1}
\{\psi_{\pm}^\mu, \psi_{\pm}^\nu\}=0\,, \qquad
\{\psi_{+}^\mu, \psi_{-}^\nu\}= \delta^{\mu \nu}\,,
\ee
as follows from the fact that both the left- and right-moving fermion
modes satisfy the Clifford algebras,
\be\label{fermanti}
\begin{array}{rcl}
{}\{\psi^\mu_r,\psi^\nu_s\} & = & \delta^{\mu\nu} \delta_{r,-s}
\\
{}\{\psi^\mu_r,\widetilde\psi^\nu_s\} & = & 0 \\
{}\{\widetilde\psi^\mu_r,\widetilde\psi^\nu_s\} & = & 
\delta^{\mu\nu} \delta_{r,-s} \,.
\end{array}
\ee
In terms of $\psi^\mu_\pm$, (\ref{fermionzero}) can be rewritten as  
\be\label{fermionzero1}
\begin{array}{lcll}
\psi^\mu_\eta |Bp,{\bf k},\eta\rangle^{(0)}_{\mbox{\tiny{R-R}}} 
& = & 0  \qquad & \hbox{$\mu=2,\ldots, p+2$} \\
\psi^\nu_{-\eta} |Bp,{\bf k},\eta\rangle^{(0)}_{\mbox{\tiny{R-R}}} 
& = & 0  \qquad & \hbox{$\nu=p+3,\ldots,9$} \,.
\end{array}
\ee
Because of the anti-commutation relations (\ref{fermanti1}) we can
define 
\be\label{eq1}
|Bp,{\bf k},+\rangle^{(0)}_{\mbox{\tiny{R-R}}} 
= \prod_{\mu=2}^{p+2} \psi^\mu_+ \prod_{\nu=p+3}^{9} \psi^\nu_- 
|Bp,{\bf k},-\rangle^{(0)}_{\mbox{\tiny{R-R}}} \,,
\ee
and then it follows that 
\be\label{eq2}
|Bp,{\bf k},-\rangle^{(0)}_{\mbox{\tiny{R-R}}} 
= \prod_{\mu=2}^{p+2} \psi^\mu_- \prod_{\nu=p+3}^{9} \psi^\nu_+ 
|Bp,{\bf k},+\rangle^{(0)}_{\mbox{\tiny{R-R}}} \,.
\ee
On the ground states the GSO-operators take the form
\be\label{GSOzerol}
(-1)^F = \prod_{\mu=2}^{9} (\sqrt{2} \psi^\mu_0) 
       = \prod_{\mu=2}^{9} (\psi^\mu_+ + \psi^\mu_-) \,,
\ee
and
\be\label{GSOzeror}
(-1)^{\widetilde{F}} 
= \prod_{\mu=2}^{9} (\sqrt{2} \widetilde{\psi}^\mu_0) 
= \prod_{\mu=2}^{9} (\psi^\mu_+ - \psi^\mu_-) \,.
\ee
Taking these equations together we then find that
\begin{eqnarray}
(-1)^F |Bp,{\bf k},\eta\rangle^{(0)}_{\mbox{\tiny{R-R}}} 
& = & |Bp,{\bf k},-\eta\rangle^{(0)}_{\mbox{\tiny{R-R}}} \\
(-1)^{\widetilde{F}} |Bp,{\bf k},\eta\rangle^{(0)}_{\mbox{\tiny{R-R}}}  
& = & (-1)^{p+1} |Bp,{\bf k},-\eta\rangle^{(0)}_{\mbox{\tiny{R-R}}} \,.
\end{eqnarray}
The action on the non-zero modes is as before, and therefore the
action of the GSO-operators on the whole boundary states is given by 
\begin{eqnarray}
(-1)^F |Bp,{\bf a},\eta\rangle_{\mbox{\tiny{R-R}}} 
& = & |Bp,{\bf a},-\eta\rangle_{\mbox{\tiny{R-R}}} \\
(-1)^{\widetilde{F}} |Bp,{\bf a},\eta\rangle_{\mbox{\tiny{R-R}}}  & = & 
(-1)^{p+1} |Bp,{\bf a},-\eta\rangle_{\mbox{\tiny{R-R}}} \,.
\end{eqnarray}

It follows from the first equation that the only potentially
GSO-invariant boundary state is of the form       
\be
\label{RR}
|Bp,{\bf a}\rangle_{\mbox{\tiny{R-R}}} = \left( 
|Bp,{\bf a},+\rangle_{\mbox{\tiny{R-R}}} 
+ |Bp,{\bf a},-\rangle_{\mbox{\tiny{R-R}}} \right)\,,
\ee
and the second equation implies that it is actually GSO-invariant if
$p$ is even (odd) in the case of Type IIA (IIB). In this case we can
find a GSO-invariant boundary state 
\be
\label{Dp}
|Dp,{\bf a}\rangle = |Bp,{\bf a}\rangle_{\mbox{\tiny{NS-NS}}} 
+ |Bp,{\bf a}\rangle_{\mbox{\tiny{R-R}}} \,.
\ee
This state satisfies world-sheet duality provided we choose
\be\label{Dpnorma}
{\cal N}^2_{\mbox{\tiny{NS-NS}}} (Dp) = {1\over 128} 
{V_{p+1} \over (2\pi)^{p+1}}   \qquad
{\cal N}^2_{\mbox{\tiny{R-R}}} (Dp) = - {1\over 8} 
{V_{p+1} \over (2\pi)^{p+1}} \,.
\ee
This gives rise to an open string consisting of a GSO-projected NS and
R sector; in particular, the GSO-projection removes the open string
tachyon from the NS sector, and the D-brane is stable. The D-brane is
also BPS since the open string spectrum is supersymmetric.

Actually the condition of world-sheet duality does not specify the
relative sign between the NS-NS and the R-R component in
(\ref{Dp}) since only the square of the normalisation constant enters
the calculation.\footnote{It also, obviously, does not specify the
overall sign, but this is just the familiar ambiguity in the
definition of quantum mechanical states.} The opposite choice of the
sign defines the anti-brane that is also BPS by itself; however, the
combination of a brane and an anti-brane breaks supersymmetry since
the two states preserve disjoint sets of supercharges. This can also
be seen from the present point of view: the open string that stretches
between a brane and an anti-brane consists of a NS and a R-sector
whose GSO-projection is opposite to that of the brane-brane or
anti-brane-anti-brane open string \cite{BS}. In particular, the open
string tachyon from the NS sector survives the projection; the system
is therefore unstable, and certainly does not preserve
supersymmetry. It is also possible to analyse the action of the
supercharges on the boundary states directly \cite{Li}. 

We have seen so far that the Type IIA (IIB) has stable BPS D-branes
for $p$ even (odd); we have also seen that the theory has unstable
D-branes for all values of $p$. However, these unstable
$\widehat{Dp}$-branes are only independent states if $p$ is odd in IIA
(and $p$ is even in IIB). In order to see this, we observe that the
normalisation of the NS-NS boundary state in (\ref{unstnorm}) is only
correct if $p$ is odd (even) in IIA (IIB). Indeed, if (\ref{unstnorm})
also held for $p$ even (odd) in IIA (IIB), the tree-diagram involving
the unstable $\widehat{Dp}$-brane and the BPS $Dp$-brane would give
rise to an open string that consists of  
\be
{1 \over \sqrt{2}} \,\left( \hbox{NS - R} \right)
\ee
and therefore violates (ii) above. The actual normalisation of
(\ref{unstnorm}) for $p$ even (odd) in IIA (IIB) is therefore
\be\label{unstnorm1}
{\cal N}^2_{\mbox{\tiny{NS-NS}}} (\widehat{Dp})= {1\over 32} 
{V_{p+1} \over (2\pi)^{p+1}} \,.
\ee
This implies that the boundary state of the unstable
$\widehat{Dp}$-brane is the sum of the boundary states of the BPS
$Dp$-brane and the BPS anti-$Dp$-brane; it therefore does not define
an additional basis vector of the lattice of D-brane states.

Finally, we should like to stress that the above analysis shows not
only that Type IIA and Type IIB has BPS $Dp$-branes for the
appropriate values of $p$, but also that these are the only 
{\em stable} D-branes of Type IIA and Type IIB. This is not
necessarily the case --- as we shall see below, some theories possess
stable D-branes that are not BPS.

\subsection{A second example: Type 0A and 0B}

As a second example let us examine the D-brane spectrum of Type 0A and
Type 0B \cite{BG1,KT,BCR,BG4}. These theories can be obtained from
Type IIA and IIB as an orbifold by $(-1)^{F^s}$, where $F^s$ is the 
spacetime fermion number. The effect of $(-1)^{F^s}$ in the untwisted
sector is to retain the bosons (\ie\ the states in the NS-NS and R-R
sectors) and to remove the fermions (\ie\ the states in the NS-R and
R-NS sectors). In the two remaining sectors, the GSO projection acts
in the usual way 
\be
\label{gsou}
\begin{array}{ll}
\hbox{NS-NS:} & P_{GSO,U}={1 \over 4} \left( 1 + (-1)^{F} \right) 
\left( 1 + (-1)^{\widetilde{F}} \right) \\
\hbox{R-R:} & P_{GSO,U}={1 \over 4} \left( 1 + (-1)^{F} \right) 
\left( 1 \pm (-1)^{\widetilde{F}} \right) \,,
\end{array}
\ee
where the $+$ sign corresponds to Type IIB, and the $-$ sign to Type 
IIA. In the twisted sector, the effect of $(-1)^{F^s}$ is to reverse
the GSO projection for both left and right-moving sectors. In addition
only the states invariant under $(-1)^{F^s}$ (\ie\ the bosons)
are retained. Thus the states in the twisted sector are again in the
NS-NS and the R-R sector, but their GSO projection is now  
\be
\label{gsot}
\begin{array}{ll}
\hbox{NS-NS:} & P_{GSO,T}={1 \over 4} \left( 1 - (-1)^{F} \right) 
\left( 1 - (-1)^{\widetilde{F}} \right) \\
\hbox{R-R:} & P_{GSO,T}={1 \over 4} \left( 1 - (-1)^{F} \right) 
\left( 1 \mp (-1)^{\widetilde{F}} \right) \,,
\end{array}
\ee
where now the $-$ sign corresponds to Type IIB, and the $+$ sign to
Type IIA. Taking (\ref{gsou}) and (\ref{gsot}) together, we can
describe the spectrum of Type 0A and Type 0B more compactly as the
subspaces of the NS-NS and R-R sectors that are invariant under the 
GSO-projection 
\be
\label{gso}
\begin{array}{ll}
\hbox{NS-NS:} & P_{GSO}={1 \over 2} 
\left(1+(-1)^{F+\widetilde{F}} \right)  
\\ 
\hbox{R-R:} & P_{GSO}={1 \over 2} 
\left(1\pm (-1)^{F+\widetilde{F}}\right) 
\,. 
\end{array}
\ee
The resulting spectrum is thus given by
\be
\begin{array}{ll}
 {\bf 0A}: & (\mbox{NS}+,\mbox{NS}+)\oplus (\mbox{NS}-,\mbox{NS}-)
       \oplus (\mbox{R}+,\mbox{R}-)\oplus (\mbox{R}-,\mbox{R}+)
       \\[5pt]
 {\bf 0B}: & (\mbox{NS}+,\mbox{NS}+)\oplus (\mbox{NS}-,\mbox{NS}-)
       \oplus (\mbox{R}+,\mbox{R}+)\oplus (\mbox{R}-,\mbox{R}-)\,.
\end{array}
\label{type0spec}
\ee
The NS-NS sector is the same for the two theories: in particular, the
low lying states consist of the ground state tachyon (that is
invariant under (\ref{gso}) since it is invariant under (\ref{gsot})),
and the bosonic part of the supergravity multiplet, \ie\ the
graviton, Kalb-Ramond 2-form, and dilaton. On the other hand, the R-R
sector is different for the two theories (as is familiar from Type
IIA and Type IIB). There are no tachyonic states, and the massless
states transform as 
\be
\begin{array}{ll}
 {\bf 0A}: \hspace{2cm} & 
({\bf 8_s} \otimes {\bf 8_c}) \oplus ({\bf 8_c} \otimes {\bf 8_s})
= 2 \cdot {\bf 8_v} + 2 \cdot {\bf 56}\,, \\[5pt]
{\bf 0B}: &
({\bf 8_s} \otimes {\bf 8_s}) \oplus ({\bf 8_c} \otimes {\bf 8_c})
= 2 \cdot {\bf 1} + 2 \cdot {\bf 28} +  {\bf 70}\,.
\end{array}
\ee
In the case of Type 0A, the theory has two 1-forms and two 3-forms in
the R-R sector, whereas Type 0B has two scalars, two 2-forms, and a
4-form (with an unrestricted 5-form field strength). The states in
the R-R sector of Type 0A and Type 0B are therefore doubled 
compared to those in Type IIA and Type IIB. One may therefore expect
that the D-brane spectrum of these theories is also doubled.

In the NS-NS sector, each boundary state $|Bp,{\bf a},\eta\rangle$ is
by itself GSO-invariant; the most general GSO-invariant boundary state
in the NS-NS sector is therefore 
\be
|Bp,{\bf a}\rangle_{\mbox{\tiny{NS-NS}}} =
\alpha_+ |Bp,{\bf a},+\rangle_{\mbox{\tiny{NS-NS}}} 
+ \alpha_- |Bp,{\bf a},-\rangle_{\mbox{\tiny{NS-NS}}} \,.
\ee
If $\alpha_+ \alpha_- \ne 0$, it follows from (\ref{+-}) that the open
string that begins and ends on the {\em same} boundary state contains
spacetime fermions. Since the closed string sector only consists of
bosons, this presumably means that the open-closed vertex of the
string field theory cannot be consistently defined; thus condition
(iii) suggests that we have to have $\alpha_+ \alpha_-=0$.\footnote{If  
the theory actually possesses one brane with $\alpha_+ \alpha_-\ne 0$,
so that the open string is NS-R with the GSO-projection
$\half(1+(-1)^F)$, the (mutually consistent) lattice of boundary
states containing this boundary state has only one stable brane (and 
anti-brane) for each allowed value of $p$; this would also seem to be
in conflict with the doubled R-R spectrum of the theory.} Thus there  
are two consistent NS-NS boundary states, and they are given by 
$|Bp,{\bf a},+\rangle_{\mbox{\tiny{NS-NS}}}$ and 
$|Bp,{\bf a},-\rangle_{\mbox{\tiny{NS-NS}}}$. As before, neither of
them is stable since the open string that begins and ends on this
state has a tachyon from the unprojected open string NS sector. In
oder to stabilise the brane, we have to add a boundary state in the
R-R sector, but as before, these are only GSO-invariant provided that
$p$ is even (odd) for Type 0A (0B). For each such $p$ we are then 
left with four different D-brane states (together with their
anti-branes) 
\be
\label{D0p}
|Dp,{\bf a},\eta,\eta'\rangle = 
|Bp,{\bf a},\eta\rangle_{\hbox{\tiny{NS-NS}}} 
+ |Bp,{\bf a}, \eta'\rangle_{\hbox{\tiny{R-R}}} \,,
\ee
where 
\be
{\cal N}^2_{\hbox{\tiny{NS-NS}}}(Dp^0) = {1\over 64} 
{V_{p+1} \over (2\pi)^{p+1}} \,, \qquad
{\cal N}^2_{\hbox{\tiny{R-R}}}(Dp^0) = - {1\over 4} 
{V_{p+1} \over (2\pi)^{p+1}} \,.
\ee
However not all of these branes are {\em mutually} consistent: the
open string between the boundary state $|Dp,{\bf a},+,+\rangle$ and 
$|Dp,{\bf b},-,+\rangle$ consists of a R-sector together with a
NS-sector with a $(-1)^F$ insertion, and likewise for 
$|Dp,{\bf a},-,-\rangle$ and $|Dp,{\bf b},+,-\rangle$. Thus there are
only two mutually consistent D-brane states for each allowed value of
$p$ which we can take to be 
\be
|Dp,{\bf a},+,+\rangle  \qquad \hbox{and} \qquad 
|Dp,{\bf a},-,-\rangle \,.
\ee
These D-branes have played an important role in recent attempts to
extend the Maldacena conjecture \cite{Maldacena} to certain
backgrounds of Type 0B string theory \cite{KT}.

\subsection{More abstract point of view: Conformal field theory with
boundaries} 

The construction of D-branes in terms of boundary states that we have
described above can be understood, from a more abstract point of view, 
as the construction of a conformal field theory on a world-sheet with
a boundary \cite{Cardy}. Given a conformal field theory that is
defined on closed world-sheets (\ie\ on closed Riemann surfaces), we
can ask the question whether we can extend the definition of this
conformal field  theory to world-sheets that have boundaries. The
prototype geometry of such a world-sheet is an infinite strip that
we take to be parametrised by $(t,s)$, where $0\leq s \leq \pi$ and
$t\in\Rop$. 

As before, we can then consider the situation where the strip is made
periodic in the $t$-direction with period $2\pi T$. The manifold is
then topologically an annulus, and the relevant partition function
becomes 
\be
Z_{\alpha\beta}(\tilde{q}) = \hbox{Tr}\; \tilde{q}^{H_{\alpha\beta}} \,,
\ee
where $\tilde{q}=e^{-2\pi T}$, $\alpha$ and $\beta$ label the boundary
conditions at either end of the strip, and $H_{\alpha\beta}$ is the
corresponding Hamiltonian. This partition function can be expressed in
terms of the representations of the chiral algebra of the conformal
field theory (see for example \cite{MRGrev} for an introduction to
these matters), 
\be\label{opentrace}
Z_{\alpha\beta}(\tilde{q}) = \sum_i n^{i}_{\alpha\beta}
\chi_i(\tilde{q}) \,,
\ee
where $\chi_i(\tilde{q})$ is the {\em character} of the representation
labelled by $i$, 
\be
\chi_i(\tilde{q}) = \tilde{q}^{-{c \over 24}} \hbox{Tr}_i \;
\tilde{q}^{L_0} \,,
\ee
and $c$ is the central charge of the corresponding Virasoro algebra. 

Under world-sheet duality, \ie\ the modular transformation 
$T\mapsto 1/T$, each character transforms as 
\be
\chi_i(\tilde{q}) = \sum_{j} S^j_i \chi_j (q) \,,
\ee
where $q=e^{-2\pi / T}$, and thus (\ref{opentrace}) becomes
\be
Z_{\alpha\beta}(\tilde{q}) = \sum_{i j} n^i_{\alpha\beta} S^j_i 
\chi_j(q)  \,.
\ee
This should then again be interpreted as the cylinder diagram between
external (boundary) states of the original bulk conformal field
theory. The closed string trace will give rise to a character of the
chiral algebra provided that each boundary state satisfies the
condition  
\be\label{standard}
\Bigl( S_{n} - (-1)^{h_S} \bar{S}_{-n} \Bigr) |B\alpha\rangle = 0 \,,
\ee
where $S$ is an arbitrary (quasi-primary) field of the chiral algebra,
and $h_S$ is its conformal weight. In particular, choosing $L=S$, we
have the condition
\be\label{conformal}
\Bigl(L_n - \bar{L}_{-n} \Bigr) |B\alpha\rangle = 0 
\ee
which is just the condition that the boundary preserves the conformal 
invariance. A solution to these conditions has been constructed by
Ishibashi \cite{Ishi} and Onogi and Ishibashi \cite{OnoIshi}, and the
corresponding coherent states are sometimes called Ishibashi
states. The actual boundary states are linear combinations of these
Ishibashi states, where the (relative) normalisations are determined by 
the condition that the numbers $n^i_{\alpha\beta}$ in
(\ref{opentrace}) are non-negative integers for all choices of
$\alpha$ and $\beta$. For left-right symmetric rational conformal
field theories (for which the chiral algebra has only finitely many
irreducible representations), explicit solutions to these constraints
have been found by Cardy \cite{Cardy}. Finally, the string field
theory condition (iii) is related to the condition that the sewing
relations of the conformal field theory are satisfied \cite{Lew}. 

For the examples of free theories (such as the bosonic Veneziano
model),\footnote{This theory is obviously not rational, and thus
Cardy's solution does not directly apply; see however \cite{HKMS}.}
the condition (\ref{standard}) for $S=\partial X^\mu$ (where $h_S=1$)
is just the condition that the boundary state represents a spacetime
spanning D-brane; the different boundary states (that are labelled by
$\alpha$ in the above) are then the different position eigenstates
(labelled by ${\bf a}$).  

In order to describe boundary states that correspond to D-branes other
than spacetime spanning D-branes, the above analysis has to be
generalised slightly. In fact, it is actually not necessary to demand
that (\ref{standard}) holds, but it is sufficient to impose
\be
\Bigl( S_{n} - (-1)^{h_S} \rho(\bar{S}_{-n}) \Bigr) |B\alpha\rangle =
0 \,, 
\ee
where $\rho$ is an automorphism of the chiral algebra that leaves the
conformal field $L$ invariant (so that (\ref{conformal}) is not
modified). With this modification, the abstract approach accounts for
all D-branes in the above model. However, it can also be generalised
to theories on curved spaces that do not possess free bosons (and for
which the definition of a Neumann or Dirichlet boundary condition is
somewhat ambiguous). In particular, this analysis has been performed
for the Gepner models \cite{RS1,GS,BDLR} and the WZW theories
\cite{PSS1,PSS2,FS,AS,ARS,FFFS}.  

It should be stressed though, that the conformal field theory analysis
that we have just sketched usually applies to the whole conformal
field theory spectrum. For theories with world-sheet supersymmetry on
the other hand, the spectrum that is relevant for string theory
consists only of a certain subspace of the conformal field theory
spectrum, namely of those states that are invariant under the
GSO-projection. Thus the conformal field theory approach has to be
slightly modified to take this into account. More significantly, the
sewing conditions of the conformal field theory only guarantee a
consistent definition of the various amplitudes for the full conformal
field theory, but it is {\it a priori} not clear whether they are
sufficient to guarantee the consistency on the GSO-invariant subspace
of string theory, \ie\ the string field theoretic consistency
conditions (see (iii) above).\footnote{This was, by the way, already
pointed out in \cite{Lew}.} At any rate, at least for the free
theories that we are considering in these lectures, most of the 
subtleties concern the nature of the GSO-projection, and therefore go
beyond at least the naive conformal field theory analysis.

\section{The non-BPS D-particle in IIB$/(-1)^{F_L} \I_4$}
\renewcommand{\theequation}{3.\arabic{equation}}
\setcounter{equation}{0}

Up to now we have described a general construction of D-branes that
does not rely on spacetime supersymmetry. We want to apply this
technique now to the construction of stable non-BPS D-branes. From the
point of view that is presented in these lectures, the simplest stable
non-BPS D-brane is presumably the D-particle of a certain orbifold of
Type IIB \cite{BG2,Sen2} (see also \cite{EP}); this shall be the topic
of this section. 

As was pointed out by Sen some time ago \cite{Sen1}, duality
symmetries in string theory sometimes predict the existence of
solitonic states which are not BPS, but are stable due to the fact
that they are the lightest states carrying a given set of charge
quantum numbers. One example Sen considered concerns the orientifold
\cite{SagCar,Hor} of Type IIB by $\Omega\I_4$ where $\I_4$ is the
inversion of four coordinates. This theory is dual to the orbifold of
Type IIB by $(-1)^{F_L} \I_4$, where $F_L$ is the left-moving
spacetime fermion number. As we shall explain below, the spectrum of
the orbifold contains in the twisted sector a massless vector
multiplet of $\N=(1,1)$ supersymmetry in $D=6$, and this implies that
the orbifold fixed-plane corresponds, in the dual orientifold theory,
to a (mirror) pair of D5-branes on top of an orientifold 5-plane
\cite{Sen1}. Because of the orientifold projection, the massless
states of the string stretching between the two D5-branes is removed,
and the gauge group is reduced from $U(2)$ to $SO(2)$. The lightest
state that is charged under the $SO(2)$ is then the first excited open
string state of the string stretching between the two D5-branes: in
the open string NS-sector the first excited states are 
\be
\begin{array}{rl}
\psi^\mu_{-3/2} |0\rangle \qquad & \hbox{8 states} \\
\alpha^\mu_{-1} \psi^\nu_{-1/2} |0\rangle \qquad & 
\hbox{$8\cdot 8=64$ states} \\
\psi^\mu_{-1/2} \psi^\nu_{-1/2} \psi^\rho_{-1/2} |0\rangle 
\qquad & 
\hbox{${8 \choose 3} = 56$ states} 
\end{array}
\ee
and in the R-sector the relevant states are 
\be
\begin{array}{rl}
\alpha^\mu_{-1} |{\bf 8_s}\rangle \qquad & 
\hbox{$8\cdot 8 = 64$ states} \\
\psi^\mu_{-1} |{\bf 8_c}\rangle \qquad & 
\hbox{$8\cdot 8 = 64$ states}\,. 
\end{array}
\ee
Thus there are altogether $128$ bosons and $128$ fermions which form 
a long (non-BPS) multiplet of the $\N=(1,1)$ $D=6$ supersymmetry
algebra. 

Since these states are stable, one should expect that the dual
(orbifold) theory also contains a stable multiplet of states that 
is charged under this $SO(2)$. However, these states are not BPS, and
the corresponding states in the dual theory therefore cannot be BPS
D-branes; in fact, as we shall show below, the orbifold theory
possesses a stable non-BPS D-particle that is stuck to the orbifold
fixed plane and that has all the above properties.

\subsection{The spectrum of the orbifold}

Let us first describe the orbifold of the Type IIB theory in some
detail. For simplicity we shall consider the uncompactified
theory, \ie\ the orbifold of $R^{9,1}/(-1)^{F_L}\I_4$. Let us choose
the convention that $\I_4$ inverts the four spatial coordinates,
$x^6,\ldots,x^9$. The fixed points under $\I_4$ form a 5-plane at
$x^6=x^7=x^8=x^9=0$, which extends along the coordinates
$x^2,\ldots,x^5$, as well as the light-cone coordinates $x^0,x^1$. In
light-cone gauge, type IIB string theory has 16 dynamical
supersymmetries and 16 kinematical supersymmetries. The former
transform under the transverse $SO(8)$ as 
\be
 Q \sim {\bf{8}}_s\;, \quad 
  \widetilde{Q}\sim {\bf{8}}_s \,.
\ee 
The orbifold breaks the transverse $SO(8)$ into 
$SO(4)_S \times SO(4)_R$, where the $SO(4)_S$ factor corresponds to 
rotations of $(x^2,\ldots,x^5)$, and the $SO(4)_R$ factor to
rotations of $(x^6,\ldots,x^9)$. The above supercharges therefore
decompose as  
\be
 \bf{8}_s \longrightarrow  
  ((\bf{2},\bf{1}),(\bf{2},\bf{1})) +
  ((\bf{1},\bf{2}),(\bf{1},\bf{2})) \,,
\ee
where we have written the representations of $SO(4)$ in terms of
$SO(4)\simeq SU(2)\times SU(2)$. The operator $\I_4$ reverses the sign
of the vector representation of $SO(4)_R$ (the $(\bf{2},\bf{2})$), and
we therefore choose its action on the $SO(4)_R$ spinors as  
\be
 \I_4 : \quad \left\{
 \begin{array}{rcr}
  (\bf{2},\bf{1}) & \rightarrow & - (\bf{2},\bf{1}) \\
  (\bf{1},\bf{2}) & \rightarrow &  (\bf{1},\bf{2}) 
 \end{array}\right.\;.
\ee
The action of $(-1)^{F_L}$ is simply
\be
 (-1)^{F_L} : \quad Q \rightarrow - Q\;, \quad
   \widetilde{Q} \rightarrow \widetilde{Q} \,,
\ee
and the surviving supersymmetries thus transform as 
\be
 Q \sim (({\bf{2}},{\bf{1}}),({\bf{2}},{\bf{1}}))\;,
 \quad
 \widetilde{Q} \sim 
    ((\bf{1},\bf{2}),(\bf{1},\bf{2})) \,.
\ee
{}From the point of view of the 5-plane world-volume this is
(dynamical, light-cone) $\N=(1,1)$ supersymmetry\footnote{The same
orbifold of type IIA would yield $\N=(2,0)$ supersymmetry.}.

The unbroken supersymmetry of these models can also be determined by 
analysing which states in the (untwisted) sector are invariant under
the orbifold projection. The NS-NS sector is the same for both IIA and
IIB, and it consists in ten dimensions of a graviton $g_{MN}$ (35
physical degrees of freedom), a Kalb-Ramond 2-form $B_{MN}$ (28) and a
dilaton $\phi$ (1). In six dimensions, the graviton gives rise to a 6d
graviton $g_{\mu\nu}$ (9), four vectors $g_{\mu i}$ (16) and ten
scalars $g_{ij}$ (10). The Kalb-Ramond 2-form gives rise to a 6d
2-form $B_{\mu\nu}$ (6), four vectors $B_{\mu i}$ (16), and six
scalars $B_{ij}$ (6). Under $\I_4 (-1)^{F_L}$ (or $\I_4$), the vectors
are all removed, and we retain a 6d graviton, a 6d 2-form and 17
scalars.  

The R-R sector of Type IIB in ten dimensions consists of a 4-form
with a self-dual 5-form field strength (35), a 2-form (28) and a
scalar (1). In six dimensions, the 4-form becomes one scalar (1), four
vectors (16) and three 2-forms (18); the 2-form becomes a 2-form (6),
four vectors (16) and six scalars (6), whilst the scalar remains a
scalar. If we orbifold by $\I_4 (-1)^{F_L}$, we retain the eight
vectors, and remove the scalars and the 2-forms; thus we have a
graviton, a 2-form, four vectors and one scalar (which combine into a
supergravity multiplet of $\N=(1,1)$) together with 4 vectors and 16
scalars (which combine into four vector multiplets of
$\N=(1,1)$).\footnote{A convenient summary of the various
supermultiplets can be found in \cite{Stra}.}  

On the other hand, if we orbifold by $\I_4$, we retain the four
2-forms and eight additional scalars. Thus we have a graviton and 5
2-forms with self-dual 3-form field strengths (that combine into a
supergravity multiplet of $\N=(2,0)$) together with 5 2-forms with
anti-self-dual 3-form field strengths and 25 scalars (which combine
into five tensor multiplets of $\N=(2,0)$).  

The analysis for Type IIA is analogous. The R-R sector in ten
dimensions consists of a 3-form (56) and a 1-form (8). In six
dimensions, the 3-form becomes seven vectors (28), four 2-forms (24)
and four scalars (4), whilst the 1-form becomes a vector (4) and four
scalars (4). If we orbifold by $\I_4 (-1)^{F_L}$, we retain the four
2-forms and the eight scalars, and therefore have the same massless
states as in the IIB orbifold by $\I_4$ giving $\N=(2,0)$
supersymmetry; if we orbifold by $\I_4$, we retain the eight vectors,
and thus obtain the same massless states as in the IIB orbifold by
$\I_4 (-1)^{F_L}$ giving $\N=(1,1)$ supersymmetry.
\smallskip
 
In addition to the untwisted sectors, the theory also contains a
twisted sector that is localised at the 5-plane. In the twisted
sector the various oscillators are moded as 
\begin{eqnarray}
 \mbox{twisted NS} : & 
  n\in  \left\{
  \begin{array}{ll}
   \bbbz & \mu=2,\ldots,5 \\
   \bbbz + 1/2 & \mu = 6,\ldots,9 
  \end{array}\right.
  \qquad 
  r\in \left\{
  \begin{array}{ll}
   \bbbz+1/2 & \mu=2,\ldots,5 \\
   \bbbz & \mu = 6,\ldots,9 
  \end{array}\right. \nonumber \\
 \mbox{twisted R} : & 
  n\in  \left\{
  \begin{array}{ll}
   \bbbz & \mu=2,\ldots,5 \\
   \bbbz +1/2 & \mu = 6,\ldots,9 
  \end{array}\right.
  \qquad 
  r\in\left\{
  \begin{array}{ll}
   \bbbz & \mu=2,\ldots,5 \\
   \bbbz + 1/2 & \mu = 6,\ldots,9 \,.
  \end{array}\right. 
\label{twisted_moding}
\end{eqnarray}
The ground state energy vanishes in both the R and NS 
sectors, and they both contain four fermionic zero modes
that transform in the vector representation of 
$SO(4)_S$ and $SO(4)_R$, respectively.
Consequently the twisted NS-NS and R-R ground states transform as  
\be
\left( ({\bf 2}, {\bf 1}) +
       ({\bf 1}, {\bf 2}) \right) \otimes
\left( ({\bf 2}, {\bf 1}) + 
       ({\bf 1}, {\bf 2}) \right) \,,
\ee
where the charges correspond to $SO(4)_S$ ($SO(4)_R$)
in the twisted R-R (NS-NS) sector.
The unique massless  representation of $D=6$ $\N=(1,1)$
supersymmetry (other than the gravity multiplet) is the 
vector multiplet 
\be
\label{vector_multiplet}
 (({\bf{2}},{\bf{2}}),({\bf{1}},{\bf{1}})) + 
 (({\bf{1}},{\bf{1}}),({\bf{2}},{\bf{2}})) +
 \mbox{fermions} \,.
\ee
In order to preserve supersymmetry, we therefore have to 
choose the GSO-projections in all twisted sectors to be of the 
form 
\be
\label{twisted_GSO}
 P_{GSO,T} = {1\over 4} \Big(1-(-1)^{F}\Big) 
              \Big(1+(-1)^{\widetilde{F}}\Big) \,.
\ee
This agrees with what we would have expected from standard 
orbifold techniques, namely that the effect of $(-1)^{F_L}$ is 
to change the left-GSO projection in the twisted sector.
In addition, the spectrum of the twisted sector must be
projected onto a subspace with either $(-1)^{F_L}\I_4=+1$
or $(-1)^{F_L}\I_4=-1$ (in the untwisted sector only $+1$
is allowed). Since twisted NS-NS (R-R) states are even (odd) 
under $(-1)^{F_L}$, and $\I_4$ reverses the sign of the vector of
$SO(4)_R$ (and leaves the vector of $SO(4)_S$ invariant), we conclude
that in the present case the twisted sector states are odd under
$(-1)^{F_L}\I_4$. 
\bigskip

Having described the spectrum and the GSO projections of the 
various sectors in some detail, we can now analyse whether a
D-brane boundary state with the appropriate properties exists. Since
the non-BPS state in the orientifold theory is localised at the
orientifold plane, one would expect that the corresponding non-BPS
D-brane should be a $\widehat{D0}$-brane that is stuck to the orbifold
fixed plane; we shall therefore analyse in the following whether such
a D-brane state exists. For definiteness we shall assume that the
$\widehat{D0}$-brane is oriented in such a way that it satisfies a
Neumann boundary condition along the $x^2$ direction.

In the (untwisted) NS-NS sector the action of $(-1)^{F_L}$ is trivial,
and $\I_4$ acts on the boundary state given in (\ref{NSNS}) as 
\be
\I_4 |Bp,{\bf a}\rangle_{\mbox{\tiny{NS-NS}}} 
= |Bp,\I_4 {\bf a}\rangle_{\mbox{\tiny{NS-NS}}} \,,
\ee
since $\I_4$ acts in the same way on left- and right-movers. If 
${\bf a}={\bf a_0}$ lies on the fixed plane, 
$\I_4 {\bf a_0}={\bf a_0}$, and the boundary state is invariant. Thus
we have a physical $p=0$ NS-NS boundary state   
\be
|U0,{\bf a_0}\rangle = |B0,{\bf a_0}\rangle_{\mbox{\tiny{NS-NS}}} \,.
\ee 
On the other hand the $p=0$ R-R boundary state is not physical since,
as we saw in section 2.2, it is not invariant under the
GSO-projection.\footnote{This boundary state is actually also not
invariant under $(-1)^{F_L}\I_4$, as follows from the analysis of
\cite{Sen2}.} 

In the twisted sector, the boundary state is of the same 
form as described before, except that now the moding of the different
fields is as described in (\ref{twisted_moding}). Since there are only
bosonic zero modes for $\mu=0,1,2,3,4,5$, and since $x^2$ is
a Neumann direction, the position of the $\widehat{D0}$-brane boundary
state is described by a 5-dimensional vector ${\bf b}$ that can be
identified with ${\bf a_0}$. Both the twisted NS-NS and the twisted
R-R sector contain fermionic zero modes, and the ground state of the 
$\widehat{D0}$-brane boundary state therefore has to satisfy 
\be
 \psi^\nu_{\eta} 
   |B0,{\bf a_0},-\eta\rangle_{\mbox{\tiny{NS-NS,T}}}^{(0)} = 0
\qquad \mbox{for $\nu=6,7,8,9$,} 
\ee
in the twisted NS-NS sector, and
\be
\begin{array}{lcll}
\psi^2_{\eta} |B0,\eta\rangle_{\mbox{\tiny{R-R,T}}}^{(0)} & = & 
             0 &  \\[5pt]
\psi^\nu_{\eta} |B0,-\eta\rangle_{\mbox{\tiny{R-R,T}}}^{(0)} & = & 
             0 & \mbox{for $\nu=3,4,5$,}
\end{array}
\ee
in the twisted R-R sector. On the ground states, the GSO operators act
as  
\be
\begin{array}{rcc}
 \mbox{twisted NS-NS :}\quad & 
  (-1)^F = \prod_{\mu=6}^{9}(\sqrt{2} \psi_0^{\mu})\,, &
  (-1)^{\widetilde{F}} = \prod_{\mu=6}^{9} 
      (\sqrt{2} \widetilde{\psi}_0^{\mu}) \\[10pt]
 \mbox{twisted R-R :}\quad &
  (-1)^F = \prod_{\mu=2}^{5} 
      (\sqrt{2} \psi_0^{\mu}) \,,&
  (-1)^{\widetilde{F}} = \prod_{\mu=2}^{5} 
      (\sqrt{2} \widetilde{\psi}_0^{\mu}) \,.
\end{array}
\ee
Using the same arguments as before in section 2.2 we then 
find  
\begin{eqnarray}
(-1)^F |B0,{\bf a_0},\eta\rangle_{\mbox{\tiny{NS-NS,T}}} & = &
|B0,{\bf a_0},-\eta\rangle_{\mbox{\tiny{NS-NS,T}}}\,, \nonumber \\
(-1)^{\widetilde{F}} 
|B0,{\bf a_0},\eta\rangle_{\mbox{\tiny{NS-NS,T}}} 
& = &   + |B0,{\bf a_0},-\eta\rangle_{\mbox{\tiny{NS-NS,T}}} \,,
\end{eqnarray}
and
\begin{eqnarray}
(-1)^F |B0,{\bf a_0},\eta\rangle_{\mbox{\tiny{R-R,T}}} & = &
|B0,{\bf a_0},-\eta\rangle_{\mbox{\tiny{R-R,T}}}\,, \nonumber \\
(-1)^{\widetilde{F}} 
|B0,{\bf a_0},\eta\rangle_{\mbox{\tiny{R-R,T}}} 
& = &   - |B0,{\bf a_0},-\eta\rangle_{\mbox{\tiny{R-R,T}}} \,. 
\end{eqnarray}
Because of (\ref{twisted_GSO}) it then follows that only the 
combination 
\be
|T0,{\bf a_0}\rangle 
= \left(|B0,{\bf a_0},+\rangle_{\mbox{\tiny{R-R,T}}}
 - |B0,{\bf a_0},-\rangle_{\mbox{\tiny{R-R,T}}} \right)\,,
\ee
in the twisted R-R sector survives the GSO-projection, and that 
no combination of twisted NS-NS sector boundary states is GSO
invariant. In addition, the ground states of the twisted R-R 
sector boundary state are odd under $(-1)^{F_L}\I_4$, as they are
precisely the vector states of $SO(4)_S$ that arise in the twisted
sector. We therefore have one further physical boundary state, and 
the total D-particle state is of the form
\be
\ket{\widehat{D0},{\bf a_0}} = \ket{U0,{\bf a_0}} + \ket{T0,{\bf a_0}}\,.
\label{D-particle}
\ee
We can then determine the cylinder diagram for a closed string that
begins and ends on the D-particle, and we find that 
\pagebreak

$$
\int_0^\infty dl \; \bra{\widehat{D0},{\bf a_0}} e^{-lH_c} 
\ket{\widehat{D0},{\bf a_0}} 
\hspace*{10cm} 
$$
\vspace*{-1.2cm} 

\be
\hspace{1cm}
= \int_0^\infty {dt\over t^{3/2}}
  \left\{ 2^{9/2}\N_{\hbox{\tiny{NS-NS}}}^2
   {f_3^8(\tilde{q}) - f_2^8(\tilde{q})\over
         f_1^8(\tilde{q})}
 + 2^{1/2} \N_{\hbox{\tiny{R-R;T}}}^2 
    {f_3^4(\tilde{q}) f_4^4(\tilde{q}) \over
          f_1^4(\tilde{q})f_2^4(\tilde{q})}\right\} \,,
\ee
where $f_i$ is defined as in (\ref{ffn}). Thus if we choose
\be
\N^2_{\hbox{\tiny{NS-NS}}}(\widehat{D0})
     = {1\over 128} {V_1 \over (2\pi)} \,, \qquad 
\N^2_{\hbox{\tiny{R-R;T}}}(\widehat{D0})
     = - \half {V_1 \over (2\pi)} \,, 
\ee
we obtain (compare \cite{Sen2}) 
\be
 \int dl\; \bra{\widehat{D0},{\bf a_0}} e^{-lH_c} 
                  \ket{\widehat{D0},{\bf a_0}} = 
 {V_1\over 2\pi} \int{dt\over 2t} \; \mbox{Tr}_{\mbox{\tiny{NS-R}}} 
   \Big[{1\over 2}(1+(-1)^F\I_4)e^{-2tH_o}\Big] \,.
\ee
The open string spectrum thus consists of a NS and a R sector, and
both are projected by $\half(1+(-1)^F\I_4)$. 
The tachyon of the NS sector is even under $\I_4$ but
odd under $(-1)^F$, and is therefore removed from the 
spectrum. This indicates that the D-particle is stable.

In addition, 4 massless states are removed from the NS sector, 
leaving 4 massless bosons, and the R sector contains 8 massless 
fermions. Including the zero modes in the light-cone 
directions,\footnote{When counting the zero modes of a D-brane
one must include the light-cone directions as well as 
the physical (transverse) massless states of the open
string. See for example \cite{Witten_bound} for a discussion
of the type IIB D-string.} this gives the D-particle 5 bosonic zero
modes and 16 fermionic zero modes. The former reflect the fact that
the D-particle is restricted to move within the 5-plane, and the
latter give rise to a long ($2^8=256$-dimensional) representation of
the six-dimensional $\N=(1,1)$ supersymmetry algebra. Finally, the
D-particle is charged under the vector field in the twisted R-R
sector. We have therefore managed to construct a boundary state that
possesses all the properties that we expected to find from the S-dual
description.  
\medskip

Sen has proposed a different realisation for this state as the ground
state of a D-string anti-D-string system \cite{Sen2}. In order to
describe this construction, it is useful to consider the theory where
at least one of the four circles that are inverted by the action of
$\I_4$ is compact. (This will serve as a preparation for the following
section where we consider the T-dual of the configuration where all
four circles are compactified.) Let us then consider a D1-brane
anti-D1-brane pair that wraps around this compact circle, $x^6$,
say. In the reduced space, the branes stretch from the fixed point at
$x^6=0$ to the fixed point at $x^6=\pi R^6$. 

As we have seen before, the ground state of the open string between
the brane and the anti-brane is a tachyon; this indicates that the
system is unstable to decay into the vacuum. However, we can consider
the configuration where we switch on a $\Zop_2$ Wilson line on either
the brane or the anti-brane. This implies that the tachyon changes
sign as we go around the circle, and thus the ground state energy
of the open string is given by 
\be
m^2 = - \half + {1\over R^2_6} \,.
\ee
In particular, the ground state of the open string is only tachyonic
if $R_6 > \sqrt{2}$; on the other hand, for $R_6<\sqrt{2}$ the ground 
state of the open string is massive, and the brane anti-brane system
is stable.  

As we shall see in the next section, the non-trivial $\Zop_2$ Wilson
line implies that the twisted R-R charge at the endpoints of the
D1-brane have opposite sign. Thus the combined system of the brane
anti-brane pair only carries twisted R-R charge at one end (but not
the other); it also does not carry any untwisted R-R charge, and
therefore has the same charges as the non-BPS D-particle that we have
just discussed (see Figure~2). 
\begin{figure}[h]
\hspace*{2.5cm}
\epsfxsize=10cm
\epsfbox{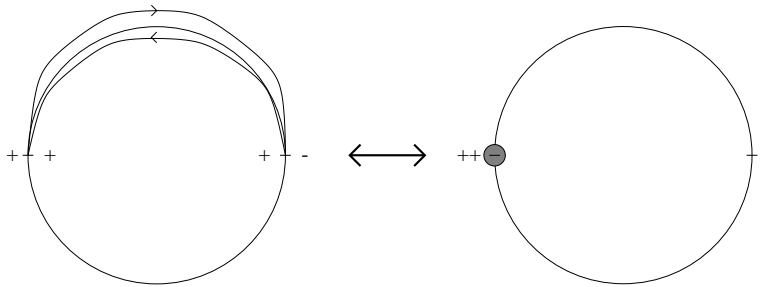}
\caption{D1-brane anti-brane pair with a relative $\Zop_2$ Wilson line
and the D0-brane.}
\end{figure}
This suggests that the brane anti-brane pair decays into the
D-particle if $R_6 > \sqrt{2}$. This interpretation is further
supported by the fact that for $R_6 < \sqrt{2}$, the open string that
begins and ends on the D-particle contains a tachyon, and thus
indicates that the D-particle is not the stable state. Indeed, the
projection $\half(1+(-1)^F\I_4)$ removes the tachyon with winding
number $0$, but the anti-symmetric combination of winding number
$w=\pm 1$ is contained in the spectrum; this state has mass 
\be
m^2 = - \half + \left({R_6\over 2}\right)^2
\ee
and thus becomes tachyonic if $R_6 < \sqrt{2}$. One can also compare
the mass and the R-R charge of the two states, and they agree indeed
(for $R_6=\sqrt{2}$).

\section{Non-BPS states in Heterotic -- Type II duality}
\renewcommand{\theequation}{4.\arabic{equation}}
\setcounter{equation}{0}

If we consider the compactification of the above IIB orbifold on a
4-torus (on which $\I_4$ acts) then the theory is T-dual to 
\be
\hbox{IIB on $T^4/ (-1)^{F_L} \I_4$}
\stackrel{\hbox{\tiny{T}}}{\longleftrightarrow}
\hbox{IIA on $T^4/ \I_4$}\,.
\ee
The orbifold of $T^4/ \I_4$ describes a special point in the moduli
space of K3 surfaces, the so-called orbifold point. On the other hand,
Type IIA on K3 is known to be S-dual to the heterotic string on $T^4$
\cite{Witten1}. 

Under T-duality, the stable non-BPS $\widehat{D0}$-brane of the Type
IIB orbifold becomes a stable non-BPS $\widehat{D1}$-brane of Type IIA
on K3; it is then natural to ask whether one can identify the
corresponding non-BPS state in the heterotic theory. This is actually
an interesting problem in its own right since both theories of the
dual pair are quantitatively under control, and one can make detailed
comparisons. The following discussion, except for section 4.3.2 that
has not been discussed before, follows closely \cite{BG3} (see also
\cite{BG5}).

\subsection{The setup}

Let us first explain the conventions of the orbifold of the Type IIA
theory. In the untwisted sector, the GSO-projections are given as in
(\ref{IIAGSO}). If we denote the compact coordinates along which
$\I_4$ acts by $x^6,\ldots,x^9$, the moding of the fields in the
twisted sectors is as in (\ref{twisted_moding}). Furthermore, the
GSO-projections in the relevant twisted sectors are given by
\begin{eqnarray}
\mbox{twisted NS-NS} & {1\over 4} \Bigl(1-(-1)^F\Bigr) 
\Bigl(1-(-1)^{\widetilde{F}}\Bigr) \,, \\
\mbox{twisted R-R} & {1\over 4} \Bigl(1-(-1)^F\Bigr)
\Bigl(1+(-1)^{\widetilde{F}}\Bigr)
\end{eqnarray}
Since the theory has  $D=6$ $\N=(1,1)$ supersymmetry, the states in
the massless R-R sector must form a vector, and thus the
GSO-projection must be the same as for the T-dual IIB/$(-1)^{F_L}\I_4$
orbifold. Consistency with the operator product expansion, in
particular the OPE
\be
\mbox{R-R} \times \mbox{R-R;T} = \mbox{NS-NS;T} 
\ee
then determines the GSO-projection in the twisted NS-NS sector; in
fact, since the GSO-projection of Type IIA and Type IIB are opposite
in the untwisted R-R sector, the same must hold in the twisted NS-NS
sector.\footnote{For a recent discussion of the subtleties associated
with the choice of the GSO-projections in the twisted sectors see
\cite{BG6}.} 

Next let us recall the precise relation between type IIA at the
orbifold point of K3 and the heterotic string on $T^4$; the following
discussion follows closely \cite{Polbook}. Let us denote the radii of
the compactified coordinates by $R_{Ai}$ and $R_{hi}$ for Type IIA and
the heterotic string, respectively. The sequence of dualities
relating the two theories is given by   
\be
 \mbox{het}\;\; T^4 \stackrel{S}{\longrightarrow}
 \mbox{I}\;\; T^4 \stackrel{T^4}{\longrightarrow}
 \mbox{IIB}\;\; T^4/\bbbz_2' \stackrel{S}{\longrightarrow}
 \mbox{IIB}\;\; T^4/\bbbz_2'' \stackrel{T}{\longrightarrow}
 \mbox{IIA}\;\; T^4/\bbbz_2 \,,
\label{sequence}
\ee
where the various $\bbbz_2$ groups are 
\be
 \bbbz_2' = (1,\Omega\I_4) \quad
 \bbbz_2'' = (1,(-1)^{F_L}\I_4) \quad
 \bbbz_2 = (1,\I_4)\,.
\ee
Here $\I_4$ reflects all four compact directions, $\Omega$ reverses
world-sheet parity, and $F_L$ is the left-moving part of the spacetime
fermion number. The first step is ten-dimensional S-duality between 
the ($SO(32)$) heterotic string and the type I string \cite{PolWit},
which relates the (ten-dimensional) couplings and radii
as\footnote{Numerical factors are omitted until the last step.} 
\be
g_I \propto g_h^{-1} \qquad \qquad 
R_{Ij} \propto g_h^{-1/2} R_{hj} \,.
\ee
The second step consists of four T-duality transformations on the
four circles, resulting in the new parameters
\be
\begin{array}{lclcl}
g' & = & V_I^{-1} g_I & \propto & V_h^{-1} g_h \\
R'_j & = & R_{Ij}^{-1} & \propto & g_h^{1/2} R_{hj}^{-1} \,,
\end{array}
\ee
where $V_I=\prod_j R_{Ij}$ and $V_h= \prod_j R_{hj}$ denote the
volumes (divided by $(2\pi)^4$) of the $T^4$ in the type I and
heterotic strings, respectively. This theory has $16$ orientifold fixed
points. In order for the dilaton to be a constant, the R-R charges have
to be canceled locally, {\it i.e.} one pair of D5-branes has to be
situated at each orientifold 5-plane. In terms of the original
heterotic theory, this means that suitable Wilson lines must be
switched on to break  $SO(32)$ (or $E_8\times E_8$) to $U(1)^{16}$;
this will be further discussed below. The third step is S-duality of
type IIB. The new parameters are given by
\be
\begin{array}{lclcl}
g'' & = & g'^{-1} & \propto & V_h g_h^{-1} \\
R''_j & = & g'^{-1/2} R'_j & \propto & V_h^{1/2} R_{hj}^{-1} \,.
\end{array}
\ee
Finally, the fourth step is T-duality along one of the compact
directions, say $x^6$. The resulting theory is type IIA on a K3 in the
orbifold limit. The coupling constants and radii are given by 
\be
\label{relation}
\begin{array}{lclcl}
g_A & = & g'' (R''_6)^{-1} & = & g_h^{-1} R_{h6} V_h^{1/2} \\
R_{Aj} & = & R''_j & = & 2 V_h^{1/2} R_{hj}^{-1} \qquad 
\mbox{for $j\ne 6$} \\
R_{A6} & = & (R''_6)^{-1} & = & 2^{-1} V_h^{-1/2} R_{h6}\,,
\end{array}
\ee
where we have now included the numerical factors (that will be shown
below to reproduce the correct masses for the BPS-states).\footnote{In
our conventions $\alpha'_h=1/2$, $\alpha'_A=1$.} In addition, the
metrics in the low energy effective theories are related as
\cite{Witten1}   
\be
  G^{A}_{\mu\nu} = V_h g_h^{-2} G^{h}_{\mu\nu} \,.
\label{metric}
\ee

The corresponding point in the moduli space of the heterotic theory
has $B=0$ and Wilson lines that can be determined in analogy with the
duality between the heterotic string on $S^1$ and type IIA on
$S^1/\Omega\I_1$ (type IA). A constant dilaton background for the
latter requires the  Wilson line $A=((\half)^8,0^8)$ in the former
\cite{ks,lowe,bgl}, resulting in the gauge group 
$SO(16)\times SO(16)$. The sixteen entries in the Wilson line
describe the positions of the D8-branes along the interval in type
IA. This suggests that the four Wilson lines in our case should be 
\begin{eqnarray}
A^9 & = & {\displaystyle \left(\left(\half\right)^8, 0^8 \right)}
  \nonumber \\
A^8 & = & {\displaystyle \left(\left(\half\right)^4,0^4,
           \left(\half\right)^4,0^4 \right)} \nonumber \\ 
A^7 & = & {\displaystyle \left(\left(\half\right)^2,0^2,
           \left(\half\right)^2,0^2,\left(\half\right)^2,0^2,
           \left(\half\right)^2,0^2 \right) }\nonumber \\
A^6 & = & {\displaystyle \left(\half,0,\half,0,\half,0,\half,0,
                \half,0,\half,0,\half,0,\half,0 \right)} \,,
\end{eqnarray}
so that there is precisely one pair of D-branes at each of the sixteen
orientifold planes. Indeed, this configuration of Wilson lines 
breaks the gauge group $SO(32)$ to $SO(2)^{16}\sim U(1)^{16}$, 
and there are no other massless gauge particles that are charged under
the Cartan subalgebra of $SO(32)$. To see this, recall that the momenta
of the compactified heterotic string are given as \cite{Ginsparg} 
\be
\begin{array}{lclcl}
{\bf P}_L & = & (P_L,p_L) & = & {\displaystyle
\left( V_K + A_K^i w_i\; ,\; {p^i \over 2 R_i}+ w^i R_i\right)}\\[10pt]
{\bf P}_R & = & \phantom{(P_R,} p_R & = & {\displaystyle
\left( {p^i \over 2 R_i} - w^i R_i  \right)\,,}
\end{array}
\ee
where $p^i$ is the physical momentum in the compact directions 
\be
 p^i = n^i + B^{ij} w_j - V^K A_K^i - {1\over 2}A_K^i A_K^j w_j \,,
\ee
$w_i,n_i\in\bbbz$ are elements of the compactification lattice
$\Gamma^{4,4}$, and $V^K$ is an element of the internal lattice
$\Gamma^{16}$. For a given momentum $({\bf P}_L,{\bf P}_R)$, a
physical state can exist provided the level matching condition    
\be
\label{level}
\half {\bf P}_L^2 + N_L - 1= \half {\bf P}_R^2 + N_R - c_R
\ee
is satisfied, where $N_L$ and $N_R$ are the left- and right-moving 
excitation numbers, and $c_R=1/2$ ($c_R=0$) for the right-moving NS
(R) sector. The state is BPS if $N_R=c_R$ \cite{dh}, and its mass is
given by  
\be
 {1\over 4}m_h^2 = \left(\half {\bf P}_L^2 + N_L - 1\right) +
                 \left(\half {\bf P}_R^2 + N_R - c_R\right) =
       {\bf P}_R^2 + 2 ( N_R - c_R)\,.
\label{heteroticmass}
\ee

The massless states of the gravity multiplet and the Cartan subalgebra
have $N_L=1$ and ${\bf P}_L={\bf P}_R=0$.  Additional massless gauge
bosons would have to have $N_L=0$, and therefore ${\bf P}_L^2 =2$.  If
$w_i=0$ for all $i$, this requires $V^2=2$ and $p_i=0$.  The possible
choices for $V$ are then simply the roots of $SO(32)$, and it is easy
to see that for each root at least one of the inner products 
$V^K A_K^i$ is half-integer; thus $p^i\in\bbbz + 1/2$ cannot vanish,
and the state is massive.  On the other hand, if $w_i\ne 0$ for at
least one $i$, the above requires $(V+Aw)^2<2$, and it follows that 
$V+Aw=0$, {\it i.e.} that the massless gauge particle is not charged
under the Cartan subalgebra of $SO(32)$.

\subsection{BPS states}

In order to test the above identification further, it is useful to
relate some of the perturbative BPS states of the heterotic string to
D-brane states in IIA on $T^4/\bbbz_2$, and to compare their masses. 
Let us start with the simplest case -- a bulk D-particle. This state
is charged only under the bulk $U(1)$ corresponding to the
ten-dimensional R-R one-form $C_{R-R}^{(1)}$. It can be described by the
boundary state  
\begin{eqnarray}
\label{D00}
|D0;{\bf a},{\bf b},\epsilon_1\rangle & = &
\Bigl( |B0;{\bf a},{\bf b}\rangle_{\hbox{\tiny{NS-NS}}} 
+ \epsilon_1 
|B0;{\bf a},{\bf b}\rangle_{\hbox{\tiny{R-R}}} \Bigr) \nonumber
\\
& & \quad + \Bigl( |B0;{\bf a},-{\bf b}\rangle_{\hbox{\tiny{NS-NS}}} 
+ \epsilon_1 |B0;{\bf a},-{\bf b}\rangle_{\hbox{\tiny{R-R}}} \Bigr)\,,
\end{eqnarray}
where ${\bf a}$ denotes the position along the uncompactified
directions for which the D-brane has Dirichlet boundary conditions,
\ie\ $x^0,x^1,x^3,x^4,x^5$, and ${\bf b}$ denotes the position along
the compacitified directions, $x^6,\ldots,x^9$. Since the directions 
$x^6,\ldots,x^9$ are compact, the corresponding momenta are quantised,
$k^i = m_i / R_{Ai}$, and the momentum integrals are replaced by sums;
thus the boundary state becomes
\be\label{1141}
|B0,{\bf a},{\bf b},\eta\rangle = {\cal N} \int 
\prod_{\nu=0,1,3,\ldots, 5} dk^\nu e^{i k^\nu a^\nu} 
\left(\prod_{i=6}^{9} \sum_{m_i\in\Zop} e^{i m_i b^i/R_{Ai}} \right)
\widehat{|B0,{\bf k},{\bf m},\eta\rangle}\,,
\ee
where $|B0,{\bf k},{\bf m},\eta\rangle$ is given by the same formula
as in (\ref{115}). The GSO-invariant boundary state 
$|B0;{\bf a},{\bf b}\rangle$ is then again given as in (\ref{NSNS}) and
(\ref{RR}). (Since we are dealing with the untwisted sector of a Type
IIA orbifold, the R-R sector boundary state with $p$ even is GSO
invariant.) The state in (\ref{D00}) is manifestly also invariant
under the orbifold operator $\I_4$ since it is the symmetric
combination of a D0-brane state together with its image under $\I_4$. 

In order to determine the correct normalisation of the different
boundary states we have to perform a similar calculation as before in
the case of the uncompactified Type IIA and Type IIB theory. There
are, however, two minor modifications. Firstly, since the momenta
along the four compact directions are quantised, one cannot simply do
the Gaussian integral; instead, one is left with a momentum sum that
can be transformed, using the Poisson resummation,
\be \label{ezzz1}
\sum_{m\in\Zop} e^{-\pi l (m/R)^2} = {R \over \sqrt{l}}
\sum_{n\in\Zop} e^{-2t\pi (nR)^2} \,,
\ee
into a winding sum which in turn appears in the open string
trace.\footnote{More details on this can be found in \cite{Sen2} and
\cite{GabSen}.} Secondly, the open string that one obtains from
(\ref{D00}) will have four sectors (depending on whether each end of
the string is at $({\bf a},{\bf b})$ or at $({\bf a},-{\bf b})$), each
of which consists of   
\be
\hbox{[NS - R]} \; {1\over 2} \Big(1+(-1)^F\Big) \,.
\ee
However, under the action of $\I_4$, the four sectors are pairwise
identified, and therefore only half as many open string states
survive. Taking this into account, the normalisation of the boundary
states in (\ref{D00}) turn out to be  
\be
R_{A6} R_{A7} R_{A8} R_{A9}
{\cal N}^2_{\hbox{\tiny{NS-NS}}} (D0) = \half {1\over 128} 
{V_{1} \over (2\pi)} \,, \qquad
R_{A6} R_{A7} R_{A8} R_{A9}
{\cal N}^2_{\hbox{\tiny{R-R}}} (D0) = - \half {1\over 8} 
{V_{1} \over (2\pi)} \,.
\ee
As before, $\epsilon_1=\pm 1$ differentiates a D-particle from an 
anti-D-particle.  

The mass of the D-particle in the Type IIA theory is given by
$m_A(D0)=1/g_A$. Using (\ref{relation}) and (\ref{metric}), the
mass of the corresponding state in the heterotic theory is therefore 
\be
m_h(D0) = V_h^{\half} g_h^{-1} m_A(D0) 
        = V_h^{\half} g_h^{-1} {1 \over g_A}
        = {1\over R_{h6}} \,.
\ee
This implies that the corresponding heterotic state has trivial
winding ($w_i=0$) and momentum ($V=0$, $p^i=0$), except for
$p_6=\epsilon_1$. Level matching then requires that $N_L=1$, and
therefore the state is a Kaluza-Klein excitation of either the
gravity multiplet or one of the vector multiplets in the Cartan
subalgebra. 
\medskip

Next we consider the D-particle that is stuck at one of the fixed
planes (which we may assume to be the fixed plane with 
${\bf b}=0$). The  mass and the bulk R-R charge of this D-particle is 
precisely half of that of the bulk D-particle that we discussed above;
it is therefore sometimes called a  `fractional' D-particle
\cite{DM}. It also carries unit charge with respect to the twisted R-R
$U(1)$ at the fixed plane. The corresponding boundary state is then
\be
\label{D0}
|D0_f,{\bf a};\epsilon_1,\epsilon_2\rangle = 
|B0;{\bf a}\rangle_{\hbox{\tiny{NS-NS}}} 
+ \epsilon_1 |B0;{\bf a}\rangle_{\hbox{\tiny{R-R}}}
+ \epsilon_2 |B0;{\bf a}\rangle_{\hbox{\tiny{NS-NS;T}}}
+ \epsilon_1 \epsilon_2 |B0;{\bf a}\rangle_{\hbox{\tiny{R-R;T}}} \,.
\ee
As we have seen above, the boundary states in the untwisted sector are
GSO- and orbifold-invariant. As regards the boundary states in the
twisted sectors, the analysis is completely analogous to the analysis
of the previous section, the only difference being that the
GSO-projection in the twisted NS-NS sector is now opposite to what it
was there; as a consequence the D0-brane boundary state is also
GSO-invariant in that sector.

The normalisation of the boundary states in the untwisted sector is as
for the case of the bulk D0-brane above, 
\be
R_{A6} R_{A7} R_{A8} R_{A9}
{\cal N}^2_{\hbox{\tiny{NS-NS}}} (D0_f) = \half {1\over 128} 
{V_{1} \over (2\pi)} \,, \qquad
R_{A6} R_{A7} R_{A8} R_{A9}
{\cal N}^2_{\hbox{\tiny{R-R}}} (D0_f) = - \half {1\over 8} 
{V_{1} \over (2\pi)} \,,
\ee
and in the twisted sectors it is 
\be\label{twisnorma}
\N^2_{\hbox{\tiny{NS-NS;T}}} (D0_f) = {1 \over 4} {V_1 \over (2\pi)}
\,, \qquad  
\N^2_{\hbox{\tiny{R-R;T}}} (D0_f) = - {1 \over 4} {V_1 \over (2\pi)}
\,. 
\ee
With these normalisations, the open string between two such
D-particles is given by  
\be 
\hbox{[NS - R]} \; {1\over 4} \Big(1+\epsilon_1\epsilon_1'(-1)^F\Big) 
\Big(1+\epsilon_2\epsilon_2'\I_4\Big)\,.  
\label{frac_D0_open}
\ee 
If we consider the limit of the bulk D0-brane state as 
${\bf b}\rightarrow 0$, \ie\ as the bulk D-particle approaches the
fixed plane, the normalisation of the boundary states of the bulk
brane is indeed {\em twice} that of the corresponding components of
the fractional brane. This demonstrates explicitly that the mass and
the untwisted R-R charge of the bulk brane is indeed twice that of the
fractional brane.  

As before, $\epsilon_1=\pm 1$ and $\epsilon_1\epsilon_2=\pm 1$
determine the sign of the bulk and the twisted charges of the
fractional brane, respectively. In the blow up of the orbifold to a
smooth K3, the fractional D-particle corresponds to a D2-brane which
wraps a supersymmetric cycle \cite{Douglas}. In the orbifold limit the
area of this cycle vanishes, but the corresponding state is not
massless, since the two-form field $B^{(2)}$ has a non-vanishing
integral around the cycle \cite{Aspinwall}. In fact $B=1/2$, and the
resulting state carries one unit of twisted charge coming from the
membrane itself, and one half unit of bulk charge coming from the
D2-brane world-volume action term 
$\int d^3\sigma \,C_{R-R}^{(1)}\wedge (F^{(2)}+B^{(2)})$.  At each 
fixed point there are four such states, corresponding to the two
possible orientations of the membrane, and the possibility of 
having $F=0$ or $F=\pm 1$ (as $F$ must be integral, the state always
has a non-vanishing bulk charge). These are the four possible
fractional D-particles of (\ref{D0}). Since there are sixteen orbifold
fixed planes, there are a total of $64$ such states.

In the heterotic string these correspond to states with internal
weight vectors of the form
\be
V=\epsilon_1 \epsilon_2 (0^{2n},1,\pm 1, 0^{14-2n}) \qquad 
(n=0,\ldots,7)\,,
\label{D0weights}
\ee
and vanishing winding and internal momentum, except for 
$p_6=\epsilon_1/2$. The sixteen twisted $U(1)$ charges in the IIA
picture correspond to symmetric and anti-symmetric combinations of the
$(2n+1)$'st and $(2n+2)$'nd Cartan $U(1)$ charges in the heterotic
picture. It follows from the heterotic mass formula
(\ref{heteroticmass}) that the mass of these states is  
\be
 m_h(D0_f) = {1\over 2R_{h6}} \,.
\ee
As before, this corresponds to the mass
\be
 m_A(D0_f) = V_h^{-1/2} \, g_h \, m_h(D0_f) 
= {1\over 2g_A} \,,
\label{fracD0}
\ee
in the orbifold of type IIA, and is thus in complete agreement with
the mass of a fractional D-particle. 

Additional BPS states are obtained by wrapping D2-branes around
non-vanishing supersymmetric 2-cycles, and by wrapping D4-branes
around the entire compact space. One can compute the mass of each of
these states, and thus find the corresponding state in the heterotic
string. Let us briefly summarise the results: 
\begin{list}{(\roman{enumi})}{\usecounter{enumi}}
\item A D2-brane that wraps the cycle $(x^i,x^j)$ where $i\ne j$ and 
$i,j\in\{7,8,9\}$ has mass $m_A = R_{Ai} R_{Aj} / (2g_A)$; in heterotic
units this corresponds to $m_h=2 R_{hk}$, where $k\in\{7,8,9\}$ is not
equal to either $i$ or $j$. The corresponding heterotic state has 
$w_k=\pm 1$, $p^l=0$, $(V \pm A_k)^2=2$, and $N_L=0$.
\item A D2-brane that wraps the cycle $(x^i,x^6)$, where $i$ is either 
$7,8$ or $9$, has mass $m_A=R_{Ai} R_{A6} / (2g_A)$; in heterotic units
this corresponds to $m_h=1/(2R_{hi})$. The corresponding heterotic
state therefore has $p^i=\pm 1/2$, $w^j=0$, $V^2=2$, and $N_L=0$.
\item A D4-brane wrapping the entire compact space has mass
$m_A = \prod_i R_{Ai} / (2g_A)$; in heterotic units this corresponds to 
$m_h= 2 R_{h6}$.  The corresponding heterotic state therefore has
$w_6=\pm 1$, $p^l=0$, $(V \pm A_4)^2=2$, and $N_L=0$.
\end{list}

\subsection{Non-BPS states}

The heterotic string also contains non-BPS states that are 
stable in certain domains of the moduli space. One should therefore
expect that these states can also be seen in the dual type IIA theory,
and that they correspond to non-BPS branes. Of course, since non-BPS
states are not protected by supersymmetry against quantum corrections
to their mass, the analysis below will only hold for $g_h\ll 1$ and
$g_A\ll 1$ in the heterotic and type IIA theory, respectively.

\subsubsection{Non-BPS D-string}

\noindent The simplest examples of this kind are the heterotic 
states with vanishing winding and momenta ($w_i=p_i=0$), and weight
vectors given by 
\be
\label{intern}
\begin{array}{lcl}
V &=& \left(0^m,\pm 2,0^{15-m}\right) \\
V' &=& \left(0^{2m},\pm 1,\pm 1, 0^{2n},\pm 1,\pm 1,0^{12-2n-2m}
    \right)\,.
\end{array}
\ee
The results of the previous section indicate that these states are 
charged under precisely two $U(1)$'s associated with two fixed points
in IIA, and are uncharged with respect to any of the other
$U(1)$'s. There are four states for each pair of $U(1)$'s, carrying
$\pm 1$ charges with respect to the two $U(1)$'s. In all cases
$V^2=4$, and we must choose $N_R=c_R+1$ to satisfy
level-matching. These states are therefore {\em not} BPS, and
transform in long multiplets of the $D=6$ ${\cal N}=(1,1)$ 
supersymmetry algebra. Their mass is given by
\be
 m_h = 2\sqrt{2}\,,
\label{het_nonbps_mass}
\ee
as follows from (\ref{heteroticmass}); in particular, the mass is
independent of the radii. 

On the other hand, these states carry the same charges as two BPS
states of the form discussed in the previous section (where the charge
with respect to the spacetime $U(1)$'s is chosen to be opposite for
the two states), and they might therefore decay into them. Whether or
not the decay is kinematically possible depends on the values of the
radii (since the masses of the BPS states are radius-dependent). In
particular, the first state in (\ref{intern}) carries the same charges
as the two BPS states with $p_6=\pm 1/2$, and weight vectors of the
form 
\begin{eqnarray}
 V_1 &=&  (0^{2n},1,1, 0^{14-2n}) \nonumber\\
 V_2 &=&  (0^{2n},1,-1, 0^{14-2n}) \,,
\label{het_decay1}
\end{eqnarray}
where we have assumed that $m$ is even and written $m=2n$; if $m$ is
odd, the two weight vectors are 
\begin{eqnarray}
 V_1 &=&  (0^{2n},1,1, 0^{14-2n}) \nonumber\\
 V_2 &=&  -(0^{2n},1,-1, 0^{14-2n}) \,,
\label{het_decay11}
\end{eqnarray}
where $m=2n+1$. The mass of each of these states is $1/(2R_{h6})$,
and the decay is therefore kinematically forbidden when  
\be 
R_{h6} < {1\over 2\sqrt{2}} \,.  
\label{het_stab1}
\ee
More generally, the above non-BPS state has the same charges as two
BPS states with $w_i=0$, and internal weight vectors
\begin{eqnarray}
V_1 & = & \pm \left(0^m,1,0^{k},1,0^{14-m-k}\right) \nonumber \\
V_2 & = & \pm \left(0^m,1,0^{k},-1,0^{14-m-k}\right) \,,
\label{het_decay2}
\end{eqnarray}
where again the non-vanishing internal momenta are chosen to be
opposite for the two states. The lightest states of this form 
have a single non-vanishing momentum, $p_i=\pm 1/2$ for one of
$i=6,7,8,9$, and their mass is $1/(2R_{hi})$. Provided that 
\be
 R_{hi} < {1\over 2\sqrt{2}} \qquad i=6,7,8,9 \,,
\label{het_stab2}
\ee
the non-BPS state cannot decay into any of these pairs of BPS states,
and it should therefore be stable. Similar statements also hold for
the non-BPS states of the second kind in (\ref{intern}).
\medskip

We should therefore expect that the IIA theory possesses a non-BPS
D-brane that has the appropriate charges and multiplicities. This
state is easily constructed: it is a non-BPS D-string that stretches
between the two fixed planes into whose fractional D-particles it can
potentially decay. Let us for simplicity consider the state that
stretches along $x^6$ from the origin to the fixed plane with
coordinates $(\pi R_{A6},0,0,0)$, and let us denote the transverse
position by ${\bf c}$ (where ${\bf c}$ has non-trivial coordinates
along $x^0,x^1,x^3,x^4,x^5$). Then the boundary state is given as 
\be
\label{D1p}
|\widehat{D1},{\bf c};\theta,\epsilon\rangle = 
 |B1,{\bf c};\theta\rangle_{\hbox{\tiny{NS-NS}}}
 + \epsilon\left( |B1,{\bf c};{\bf 0}\rangle_{\hbox{\tiny{R-R;T}}} 
   + e^{i\theta} 
     |B1,{\bf c};(\pi R_{A6},0,0,0)\rangle_{\hbox{\tiny{R-R;T}}}
\right)  \,,
\ee
where $\theta$ denotes a Wilson line which originates from the fact
that the $x^6$ direction is compact so that the NS-NS vacuum can be
characterised by a winding number $w_6$.\footnote{The relevant closed
string Hamiltonian contains then also an additional term 
$v^2/(4\pi)$, where $v$ is the winding length.} In fact, the
boundary state $|B1,{\bf c};\theta\rangle_{\hbox{\tiny{NS-NS}}}$ is
defined by  
\be
|B1,{\bf c};\theta\rangle_{\hbox{\tiny{NS-NS}}} = \sum_{w_6}
 e^{i\theta w_6} |B1,{\bf c};w_6\rangle_{\hbox{\tiny{NS-NS}}} \,,
\ee
where $|B1,{\bf c};w_6\rangle_{\hbox{\tiny{NS-NS}}}$ is given by
(\ref{NSNS}), (\ref{1141}) and (\ref{115}) except that 
$|B1,{\bf k},\eta\rangle^{(0)}$ in (\ref{115}) is now replaced by  
\be
|B1,{\bf k},w_6,\eta \rangle^{(0)} \,.
\ee
This tachyonic ground state has winding number $w_6$ along the $x^6$
direction, and momentum equal to $k^i$ for $i\ne 6$. Because it
describes a $\widehat{D1}$-brane with a Neumann direction along $x^2$,
we also have that $k^2=0$; furthermore the momenta $k^i$ for $i=7,8,9$
are again quantised. This boundary state is (as before) obviously 
invariant under the GSO-projection; invariance under the orbifold
projection requires that $\theta=0$ or $\theta=\pi$ (since $\I_4$ maps
$w_6\mapsto - w_6$.) The correct normalisation will turn out to be  
\be\label{NSNSnorma}
R_{A7} R_{A8} R_{A9} 
{\cal N}^2_{\hbox{\tiny{NS-NS}}} (\widehat{D1}) = {1\over 64} 
{V_{2} \over (2\pi)^2} \,,
\ee
where $V_{2} = \pi R_{A6} V_{1}$, with $V_{1}$ being the volume along
the $x^2$-direction along which the D1-brane has a Neumann boundary
condition.  

The two boundary states in the twisted R-R sector are localised at
different fixed planes, and are otherwise standard boundary
states. Since the twisted R-R sector does not have any fermionic zero
modes in the $x^6$ direction, the ground state satisfies the same zero
mode condition as the D0-brane boundary state discussed above; this
also implies that it is GSO-invariant. The parameter $\epsilon$ takes
the values $\pm 1$, and determines the sign of the twisted R-R charge
at one end of the $\widehat{D1}$-brane. The correct normalisation will
turn out to 
be 
\be\label{RRTnorma}
{\cal N}^2_{\hbox{\tiny{R-R;T}}} (\widehat{D1}) = - {1\over 4} 
{V_{1} \over (2\pi)} \,,
\ee
where $V_1$ is the world-volume along the $x^2$ direction.

In order to describe the corresponding open string it is convenient to
use a different description for the orbifold \cite{Sen2}. Let us
denote, as before, by $\I_4$ the reflection of the four coordinates 
$x^6,\ldots,x^9$, and let $\I_4'$ be defined by 
\be
\I_4' : \left\{
\begin{array}{ll}
x^i \mapsto - x^i & \hbox{if $i\ne 6$} \,, \\
x^6 \mapsto 2 \pi R_{A6} - x^6\,. & 
\end{array}
\right.
\ee
Let us consider the compactification where initially the radius of the
sixth circle is $2 R_{A6}$. The $\Zop_2\times\Zop_2$ orbifold
of this theory that is generated by $\I_4$ and $\I_4'$ is then
equivalent to the above orbifold. In order to see this we observe that 
$\I_4$ and $\I_4'$ commute with each other, and that both are of order
two. The $\Zop_2\times\Zop_2$ orbifold can therefore equivalently be
described as the $\I_4$-orbifold of the $\I_4'\I_4$-orbifold; however, 
$\I_4\I_4'$ is the translation $x^6\mapsto x^6+2\pi R_{A6}$, and its
effect is simply to reduce the radius from $2R_{A6}$ to $R_{A6}$. 

For the above choice of normalisation constants, the spectrum of open
strings that begin and end on the above D-string is then given as 
\be
\hbox{[NS - R ]} \; {1 \over 4} \left( 1+ (-1)^F \I_4 \right) 
          \left( 1 + (-1)^F \I'_4 \right) \,,
\label{open_D1}
\ee
where the terms that involve $\I_4$ come from the twisted R-R sector
localised at ${\bf 0}$, the terms involving $\I_4'$ come from the
twisted R-R sector localised at $(\pi R_{A6},0,0,0)$, and the
remaining terms arise from the untwisted NS-NS sector. (More
specifically, the term with the unit operator corresponds to the
contribution where $w_6$ is even, whereas the term 
$1/4 (-1)^F \I_4 (-1)^F \I'_4 = 1/4 (x^6\mapsto x^6 + 2 \pi R_{A6})$
comes from the terms with $w_6$ odd.)

Since $\theta$ and $\epsilon$ can only take two different values each, 
there are four different D-strings for each pair of orbifold
points. These four D-strings are only charged under the two twisted
sector $U(1)$s associated to the two fixed planes, and the four
different configurations correspond to the four different sign
combinations for the two charges. These charges are of the same
magnitude as those of the  fractional D-particles, since the ground
state of twisted R-R sector contribution satisfies the same zero-mode
condition, and has the same normalisation (compare (\ref{twisnorma})
and (\ref{RRTnorma})). Furthermore, it follows from (\ref{open_D1})
that the D-strings have sixteen (rather than eight) fermionic zero
modes, and therefore transform in long multiplets of the $D=6$, 
${\cal N}=(1,1)$ supersymmetry algebra. These states therefore have
exactly the correct properties to correspond to the above non-BPS
states of the heterotic theory.

The open string NS sector in (\ref{open_D1}) contains a 
tachyon. However, since the tachyon is $(-1)^F$-odd, and since $\I_4$
reverses the sign of the momentum along the D-string, the
zero-momentum component of the tachyon field on the D-string is
projected out. Furthermore, since $\I_4\I'_4$ acts as 
$x^6\rightarrow x^6 + 2\pi R_{A6}$, the half-odd-integer momentum 
components are also removed, leaving a lowest mode of unit momentum.
As a consequence, the mass of the tachyon is shifted to 
\be
 m^2_T =  - \half + {1\over R_{A6}^2} \,.
\ee
For $R_{A6}<\sqrt{2}$ the tachyon is actually massive, and thus
the non-BPS $\widehat{D1}$-brane is stable. On the other hand, for 
$R_{A6}>\sqrt{2}$ the configuration is unstable and decays into the
configuration of two D-particles that sit at either end of the
interval. These D-particles carry opposite untwisted R-R charge, and
their twisted R-R charge is determined in terms of the twisted R-R
charge of the D-string at either end. 

\begin{figure}[h]
\hspace*{2.5cm}
\epsfxsize=10cm
\epsfbox{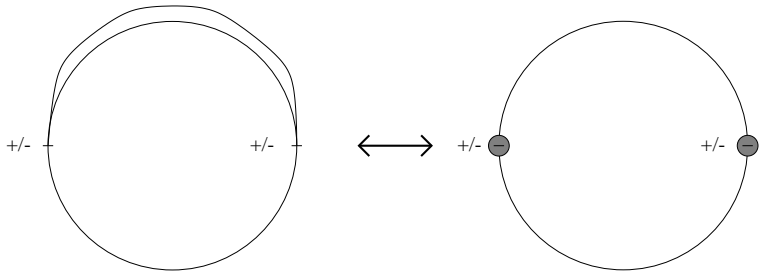}
\caption{The non-BPS $\widehat{D1}$-brane and the two fractional
D0-branes into which it can decay.}
\end{figure}

One can also understand this instability from the point of view of the
two fractional BPS D-particles. Since they carry opposite untwisted
R-R charge, the open string between them consists of 
\be 
\hbox{[NS - R]} \; {1\over 4} \Big(1-(-1)^F\Big)
\Big(1 \pm \I_4\Big)\,.  
\ee 
The ground state of the open string NS sector therefore has a mass 
\be
m^2 = -\half + \left(\pi R_{A6} T_0\right)^2
  = -\half + \left({R_{A6}\over 2}\right)^2\,,
\ee
and so becomes tachyonic for $R_{A6}<\sqrt{2}$, indicating an
instability to decay into the non-BPS D-string. The D-string can
therefore be thought of as a bound state of two fractional BPS
D-particles located at different fixed planes. This is also confirmed
by the fact that the classical mass of the D-string (\ref{D1mass}) is
smaller than that of two fractional  
D-particles (\ref{fracD0}) when 
\be 
 R_{A6}< \sqrt{2} \,,
\label{IIA_stab1}
\ee
and thus the D-string is stable against decay into two fractional
D-particles in this regime (see Figure~3). In terms of the heterotic
string, this decay channel corresponds to (\ref{het_decay1}). Given
the duality relation (\ref{relation}), the domain of stability of the
non-BPS D-string (\ref{IIA_stab1}) becomes in terms of the heterotic
moduli 
\be
V_h^{-1/2} R_{h6} < 2 \sqrt{2} \,.
\ee
Thus the D-string is stable provided that $R_{h6}$ is sufficiently
small; this agrees qualitatively with the regime of stability in the 
heterotic theory (\ref{het_stab1}). (Since we are dealing with non-BPS
states, one should not expect that these regimes of stability match
precisely.) 

Other decay channels become available to the D-string when the other
distances $R_{Ai}$ ($i=7,8,9$) become small. In particular, the
D-string along $x^6$ can decay into a pair of 
D2-branes carrying opposite bulk charges, \ie\ a D2-brane and
an anti-D2-brane, that wrap the $(x^i,x^6)$ cycle.

\begin{figure}[h]
\hspace*{2.5cm}
\epsfxsize=10cm
\epsfbox{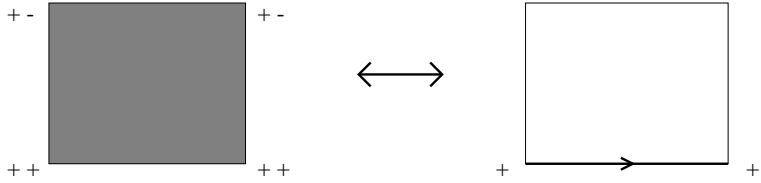}
\caption{A D2-brane anti-brane pair and the non-BPS
$\widehat{D1}$-brane into which it can decay. The twisted R-R charge
of each D2-brane at each of the four corners is one half of that of
the non-BPS $\widehat{D1}$-brane.} 
\end{figure}

Since the mass of each D2-brane in the orbifold metric is
$R_{Ai}R_{A6}/(2g_A)$, the D-string is stable in this channel when
\be
 R_{Ai} > {1 \over \sqrt{2}} \quad (i=7,8,9)\,.
\label{IIA_stab2}
\ee
The D-string can therefore also be thought of as a bound state of
two BPS D2-branes. This decay channel can also be understood from the
appearance of a tachyon on the D-string carrying one unit of winding
in the $x^i$ direction, when $R_{Ai}<1/\sqrt{2}$ \cite{Sen5}, 
or alternatively from the appearance of a tachyon between the
two D2-branes when $R_{Ai}>1/\sqrt{2}$. In terms of the heterotic
string, these decay channels are described by
(\ref{het_decay2}). Using the duality relation (\ref{relation}) as
before, (\ref{IIA_stab2}) then becomes 
\be
V_h^{-1/2} R_{hj} < 2 \sqrt{2} \qquad \hbox{for $j\ne 6$}\,.
\ee
Thus the D-string is stable against this decay provided that $R_{hj}$
is sufficiently small, and this agrees again qualitatively with the
heterotic domain of stability (\ref{het_stab2}). A similar analysis
can also be performed for D-strings that stretch between any two fixed
points.    
\smallskip

One can also compare the mass of the non-BPS $\widehat{D1}$-brane with
that of the dual heterotic state. As we mentioned before, one should
not expect that these are related exactly by the duality map since for
non-BPS states the masses are not protected from quantum
corrections. Indeed, the classical mass of the above non-BPS
$\widehat{D1}$-brane is given 
by 
\be
 m_A(\widehat{D1}) = {R_{A6}\over \sqrt{2} g_A} \,,
\label{D1mass}
\ee
where the factor of $\sqrt{2}$ comes from the fact that the
normalisation of the untwisted NS-NS component (\ref{NSNSnorma}) is by
a factor of $\sqrt{2}$ larger than that of the standard BPS D-brane of
Type II (\ref{Dpnorma}). In heterotic units, this mass is 
$\propto 1/V_h$, and therefore does not agree with
(\ref{het_nonbps_mass}).   

In the blow up of the orbifold to a smooth K3, the non-BPS D-strings
correspond to membranes wrapping pairs of shrinking 2-cycles. Since
such curves do not have holomorphic representatives, the states are
non-BPS. For each pair of 2-cycles there are four states, associated
with the different orientations of the membrane; the
membrane can wrap both cycles with the same orientation, or with
opposite orientation.  In either case the net bulk charge due to
$B=1/2$ can be made to vanish by turning on an appropriate
world-volume gauge field strength ($F=\pm 1$ in the first case, and
$F=0$ in the second).  The decay of the non-BPS D-string into a pair
of fractional BPS D-particles is described in this picture as the
decay of this membrane into two separate membranes, that wrap
individually around the two 2-cycles. It would be interesting to
understand in more detail how non-BPS branes behave away from the
orbifold point; first steps in this direction have recently been taken
in \cite{MajSen}.
\medskip

Finally, the entire discussion also has a parallel in the T-dual
theory that we analysed in the previous section. The non-BPS D-string
that stretches along $x^6$ is mapped under T-duality to the non-BPS
D-particle of the IIB orbifold. (The two different values
$\theta=0,\pi$ correspond to the two possible positions, and
$\epsilon$ to the sign of the charge of the D-particle.) The non-BPS
D-string can be formed as a bound state of a fractional D-particle and
a fractional anti-D-particle (see Figure~3). Under T-duality, the
D-particle anti-D-particle pair becomes a pair of a  BPS D-string and
an anti-D-string of the IIB theory that stretch along the $x^6$
direction; since the D-particles sit on different fixed planes, the 
BPS D-strings have a relative Wilson line. Thus the non-BPS D-particle
can be understood as the bound state of a D1-brane anti-D1-brane pair
with a relative Wilson line; this reproduces precisely the
construction of Sen \cite{Sen2}. By T-duality it follows that the
D-particle is stable provided that\footnote{ In the previous section
we considered the uncompactified theory where all radii are infinite;
in this regime the D-particle is stable.} 
\be
R_{i} \geq {1\over\sqrt{2}} \qquad \hbox{$i=6,7,8,9$.}
\ee
Similarly the other decay channels can also be related to decay
channels considered by Sen.

\subsubsection{Non-BPS $\widehat{D3}$-brane}

In addition to the non-BPS $\widehat{D1}$-brane, the IIA theory also
has a non-BPS $\widehat{D3}$-brane for which an analogous analysis
applies. The corresponding boundary state has a component in the
untwisted NS-NS sector, and a component in each of the eight twisted
R-R sectors that are localised at the vertices of the cube along which
the $\widehat{D3}$-brane stretches. The $\widehat{D3}$-brane is
characterised by three $\Zop_2$ Wilson lines (that determine the
relative signs of the twisted R-R charge at the different end-points),
and one additional sign (that determines the overall sign of the
twisted R-R charges). The states in the twisted R-R sector are again
GSO-invariant, since their ground state satisfies the same fermionic
zero mode conditions as the D0-brane state. Furthermore, a careful
analysis of the boundary state reveals \cite{GabSte} that the non-BPS
$\widehat{D3}$-brane carries at each corner precisely one half of the
twisted R-R sector charge of a fractional D0-brane. This normalisation
is consistent with the decay process of the non-BPS
$\widehat{D3}$-brane into a D2-brane anti-brane pair (see Figure~5) 
that is the analogue of the decay process of
Figure~3.\footnote{Indeed, the decay process of Figure~3 implies that 
the twisted R-R charge of each end of a non-BPS $\widehat{D1}$-brane
is the same as that of a BPS D0-brane, and the decay process of
Figure~4 implies that this charge is twice as large as the twisted R-R
charge of a BPS D2-brane at each of its four corners.}  
\begin{figure}[h]
\hspace*{2.5cm}
\epsfxsize=10cm
\epsfbox{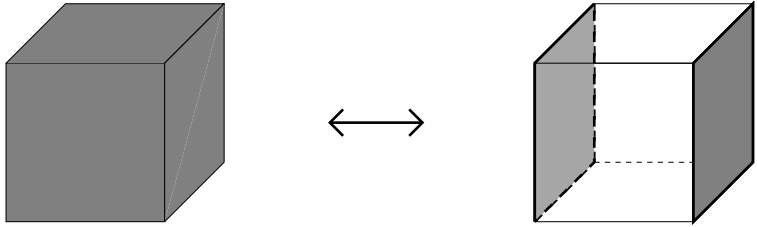}
\caption{A non-BPS $\widehat{D3}$-brane and the D2-brane anti-brane
pair into which it can decay.}
\end{figure}
The non-BPS $\widehat{D3}$-brane is stable against this decay provided
that the three radii along which it stretches are each smaller than
$\sqrt{2}$. There is also a decay channel along which the
$\widehat{D3}$-brane can decay into a D4-brane anti-D4-brane pair
(this is the analogue of the decay process of Figure~4), and the
non-BPS $\widehat{D3}$-brane is stable against this decay process
provided that the transverse radius is larger than $1/\sqrt{2}$. 

In order to identify the corresponding states in the heterotic string
it is convenient to consider the different non-BPS
$\widehat{D3}$-brane states (that are characterised by four signs and
their position in the $T^4$) in conjunction with those non-BPS
$\widehat{D3}$-brane states that correspond to the configuration where
a non-BPS $\widehat{D1}$-brane is embedded within the
$\widehat{D3}$-brane.\footnote{One can presumably describe this
configuration also as a non-BPS $\widehat{D3}$-brane with a
non-trivial magnetic flux. It would be interesting to understand this
in more detail.} Since the magnitude of the twisted R-R charge at the
end-point of the non-BPS $\widehat{D1}$-brane is twice that of the
non-BPS $\widehat{D3}$-brane, the sign of the twisted R-R charge of
the bound state differs at two vertices from that of the original 
$\widehat{D3}$-brane. Proceeding in this way, we can change the signs
of the charges at an even number of endpoints, and thus obtain
$\widehat{D3}$-brane states with $2^7=128$ different sign combinations
at the eight end-points. (For conventional $\widehat{D3}$-branes, the
number of combinations was only $2^4=16$.) In addition we can localise
the $\widehat{D3}$-brane in $2\cdot 15=30$ different ways: there are
fifteen different direction vectors between the vertices of the unit
cell, and we can choose the $\widehat{D3}$-brane to be orthogonal to
any one of them; for each such orientation, we can then localise the
$\widehat{D3}$-brane at two different positions. Taking all of this
together we are therefore looking for $30\cdot 2^7$ states in the
heterotic theory.  

The states that correspond to these non-BPS D3-branes in the dual
heterotic theory must be charged under eight of the sixteen $U(1)$s
that are described following (\ref{D0weights}), but not under any of
the other $U(1)$s. The charge with respect to each of these eight
$U(1)$s must be precisely half of that of the states in
(\ref{D0weights}). Furthermore, for each allowed set of eight such
$U(1)$s (there are $30$ such sets that correspond to the different
localisations of the D3-brane) there are $128$ such states that differ
by the signs of the charges at the end points. Heterotic states with
these properties can be found as follows: there are $128$ states that
are only charged under the first eight $U(1)$s, and the corresponding
internal weight vectors are of the form 
\be\label{D31}
\Bigl({\bf a_1}, {\bf a_2}, {\bf a_3}, {\bf a_4},0^8\Bigr)\,,
\ee
where ${\bf a_i}$ is a two-dimensional vector which is equal to one of
the following four vectors
\be
{\bf e_1}=(1,0) \,, \qquad {\bf e_2}=(-1,0)\,, \qquad 
{\bf f_1}=(0,1) \,, \qquad {\bf f_2}=(0,-1)\,.
\ee
Of the $4^4=256$ combinations only those are allowed (\ie\ have
integer inner product with $A^6$) where an even number of the 
${\bf a_i}$ are equal to ${\bf e_1}$ or ${\bf e_2}$ (and an even
number of the ${\bf a_i}$ are equal to ${\bf f_1}$ or ${\bf f_2}$);
this reduces the number of possibilities by half to the desired
$128$. It is not difficult to check that all of these states are only
charged under the first eight $U(1)$s (provided we choose the momentum
and winding numbers appropriately), and that the magnitude of the
corresponding charge is precisely half that of the states in
(\ref{D0weights}). Furthermore, these are the only states with this
property. 

We can similarly construct states that are charged under eight $U(1)$s
by choosing different positions for the four ${\bf a_i}$ vectors in
the sixteen dimensional space. Since the resulting states should not
be charged under any other $U(1)$s, we have to demand that the
internal weight vectors have integer inner product with all four
Wilson lines; the possible configurations are then
\be
\begin{array}{ll}
\Bigl({\bf a}, {\bf a}, {\bf a}, {\bf a}, {\bf 0}, {\bf 0}, {\bf 0},
{\bf 0}\Bigr)\,, \qquad & 
\Bigl({\bf 0}, {\bf 0}, {\bf 0}, {\bf 0}, {\bf a}, {\bf a}, {\bf a},
{\bf a}\Bigr)\,, \\[5pt]
\Bigl({\bf a}, {\bf a}, {\bf 0}, {\bf 0}, {\bf a}, {\bf a}, {\bf 0},
{\bf 0}\Bigr)\,, \qquad & 
\Bigl({\bf 0}, {\bf 0}, {\bf a}, {\bf a}, {\bf 0}, {\bf 0}, {\bf a},
{\bf a}\Bigr)\,, \\[5pt]
\Bigl({\bf a}, {\bf a}, {\bf 0}, {\bf 0}, {\bf 0}, {\bf 0}, {\bf a},
{\bf a}\Bigr)\,, \qquad & 
\Bigl({\bf 0}, {\bf 0}, {\bf a}, {\bf a}, {\bf a}, {\bf a}, {\bf 0},
{\bf 0}\Bigr)\,, \\[5pt]
\Bigl({\bf a}, {\bf 0}, {\bf a}, {\bf 0}, {\bf a}, {\bf 0}, {\bf a},
{\bf 0}\Bigr)\,, \qquad & 
\Bigl({\bf 0}, {\bf a}, {\bf 0}, {\bf a}, {\bf 0}, {\bf a}, {\bf 0},
{\bf a}\Bigr)\,, \\[5pt]
\Bigl({\bf a}, {\bf 0}, {\bf a}, {\bf 0}, {\bf 0}, {\bf a}, {\bf 0},
{\bf a}\Bigr)\,, \qquad & 
\Bigl({\bf 0}, {\bf a}, {\bf 0}, {\bf a}, {\bf a}, {\bf 0}, {\bf a},
{\bf 0}\Bigr)\,, \\[5pt]
\Bigl({\bf a}, {\bf 0}, {\bf 0}, {\bf a}, {\bf a}, {\bf 0}, {\bf 0},
{\bf a}\Bigr)\,, \qquad & 
\Bigl({\bf 0}, {\bf a}, {\bf a}, {\bf 0}, {\bf 0}, {\bf a}, {\bf a},
{\bf 0}\Bigr)\,, \\[5pt]
\Bigl({\bf a}, {\bf 0}, {\bf 0}, {\bf a}, {\bf 0}, {\bf a}, {\bf a},
{\bf 0}\Bigr)\,, \qquad & 
\Bigl({\bf 0}, {\bf a}, {\bf a}, {\bf 0}, {\bf a}, {\bf 0}, {\bf 0},
{\bf a}\Bigr)\,.
\end{array}
\ee
There are fourteen different such classes of states, and this
construction accounts therefore for $14\cdot 128$ states. 

The remaining $16\cdot 128$ states correspond to states in the 
{\em spinor representation} of $SO(32)$. These are the states whose
internal weight vectors are of the form
\be
\Biggl( \pm \half, \pm \half, \pm \half, \pm \half, 
       \pm \half, \pm \half, \pm \half, \pm \half,
       \pm \half, \pm \half, \pm \half, \pm \half,
       \pm \half, \pm \half, \pm \half, \pm \half \Biggr) \,,
\ee
where the number of $+$ signs is even. Each of these $2^{15}$ states
is charged under eight of the sixteen internal $U(1)$s. In order for
the state to be uncharged under any other $U(1)$, we have to demand
again that the inner product of the internal weight vector with each
of the four Wilson lines is integral. For each Wilson line, this
condition selects one half of the states, and since the four
conditions are independent of each other, the number of states that
have this property for all four Wilson lines is $2^{11}=16\cdot 128$. 
Together with the above $14\cdot 128$ states we have therefore found
all $30\cdot 128$ states that correspond to $\widehat{D3}$-branes
(including those that contain $\widehat{D1}$-branes within). It is
also easy to see that these are all the states in the heterotic theory
that have the above properties! 

As we have seen above, there are $30\cdot 16$ conventional non-BPS
$\widehat{D3}$-brane configurations; these are mapped under T-duality
(of all four circles) to the various non-BPS $\widehat{D1}$-brane
configurations that we have discussed before; their number is  
\be
4 \cdot {16 \choose 2} = 30 \cdot 16
\ee
and is therefore in agreement with the above. The remaining bound
states of non-BPS $\widehat{D3}$-branes with non-BPS
$\widehat{D1}$-branes are mapped into themselves under T-duality. 

One can also analyse the stability of these non-BPS states in both
theories. For example, the spinor state with internal weight vector
\be\label{spinor}
\Biggl(- \half, \half, \half, \half, \half, \half, \half, -\half,
   -\half, \half, \half, \half, \half, \half, \half, -\half \Biggr)
\ee
has the same charges as the two BPS states with momenta
\be\label{D3dec}
\begin{array}{lcl}
(P_L,p_L;p_R)_1 & = & 
\Bigl(-\half,(\half)^6,-\half,0^8),R_{h9};-R_{h9} \Bigr) \\
(P_L,p_L;p_R)_2 & = & 
\Bigl((0^8, -\half,(\half)^6,-\half),-R_{h9};R_{h9} \Bigr) \,.
\end{array}
\ee
The mass of all of the above non-BPS states in the heterotic theory is
$m_h=2\sqrt{2}$, whereas the mass of each of the two BPS states in 
(\ref{D3dec}) is $m_h=2 R_{h9}$; thus the non-BPS state (\ref{spinor})
is stable against the decay into (\ref{D3dec}) provided that 
\be\label{test}
R_{h9}> {1\over \sqrt{2}} \,.
\ee

The two BPS states in (\ref{D3dec}) correspond, in the IIA theory, to 
two D2-branes that extend along the $x^7,x^8$ plane (this follows
from the analysis at the end of section~4.2.), and the non-BPS
$\widehat{D3}$-brane extends along the $x^7,x^8,x^9$ directions. The
decay process that we are considering is therefore that depicted in 
Figure~5. The mass of the $\widehat{D3}$-brane is 
\be
m_A (\widehat{D3}) = {1 \over \sqrt{2} g_A} R_{A7} R_{A8} R_{A9} \,, 
\ee
whereas the mass of each of the two D2-branes is 
\be
m_A(D2) = {1 \over 2 g_A} R_{A7} R_{A8} \,.
\ee
The non-BPS $\widehat{D3}$-brane is therefore stable against this
decay process provided that 
\be
R_{A9} < \sqrt{2} \,.
\ee
(This was, by the way, already mentioned following Figure~5.) In terms
of the heterotic theory the last equation becomes 
\be
V_{h}^{-\half} R_{h9} > \sqrt{2}\,.
\ee
Again, this agrees qualitatively with (\ref{test}). The other cases
are similar.

\subsection{Bose-Fermi degeneracy}

BPS D-branes carrying identical charges do not exert any force on each
other, and can be at equilibrium at all distances. This is a
consequence of supersymmetry, and reflects the fact that the spectrum
of open strings living on the world volume of the system has exact
degeneracy between bosonic and fermionic states at all mass levels. As
a result the partition function of open strings, which corresponds
to the negative of the interaction energy of the pair of D-branes,
vanishes identically. 

A non-BPS D-brane (such as the D-branes we have analysed above) breaks
supersymmetry and the spectrum of open strings that begin and end on
it does in general not have exact Bose-Fermi degeneracy. The open
string partition function, and hence the interaction energy of a pair
of such D-branes, is then not zero. The D-branes then exert a force on
each other, and the system is not in equilibrium.    

It was observed in \cite{GabSen} that the partition function depends
non-trivially on the moduli (in particular the four radii), and that
there exist special points in the moduli space where the spectrum
develops exact Bose-Fermi degeneracy. For definiteness let us consider
the case of the non-BPS D-particle of the IIB orbifold. We are
interested in the situation where all four directions along which the
orbifold acts are compact; the boundary state for the D-particle is
then given as in the previous section, except that the momentum
integrals along $x^6\ldots,x^9$ are replaced by sums, and that the
normalisation constant in the untwisted NS-NS sector is changed to 
\be
R_{6} R_{7} R_{8} R_{9} \N^2_{\hbox{\tiny{NS-NS}}} (\widehat{D0}) =
{1\over 128} {V_1 \over (2\pi)} \,.
\ee
(Details of this can again be found in \cite{GabSen}.) The open string
partition function is then given by 
\be \label{ex10}
Z = {1\over 2} \int {dt\over 2t} {V_1\over (2\pi)}
(2t)^{-{1\over 2}}
\left[ {f_4(\tilde q)^8\over f_1(\tilde q)^8}\left(\prod_{i=6}^9
\sum_{n_i\in \Zop} \tilde q^{2R_i^2 n_i^2}\right)
- 4 \cdot {f_3(\tilde q)^4 f_4(\tilde q)^4\over f_1
(\tilde q)^4 f_2(\tilde q)^4}\right]\, .
\ee
Let us now consider the critical case where $R_i={1\over \sqrt 2}$ for
each $i=6,7,8,9$. In this case we get
\be \label{ex11}
\sum_{n_i\in \Zop} \tilde q^{2R_i^2 n_i^2} =
\sum_{n\in \Zop} \tilde q^{n^2}\, .
\ee
Using the sum and the product representation of the Jacobi
$\vartheta$-function $\vartheta_3(0|\tau)$ \cite{EMOT53},
\be \label{ex12}
\vartheta_3(0|\tau) = \sum_{n\in \Zop} \tilde q^{n^2}
= \prod_{n=1}^\infty (1 -
\tilde q^{2n}) (1 + \tilde q^{2n-1})^2
= f_1(\tilde q) f_3^2(\tilde q)\, ,
\ee
where $\tilde q=e^{2\pi i\tau}$, and the identity
\be \label{ex13}
f_4(\tilde q) {1\over \sqrt 2} f_2(\tilde q) f_3(\tilde q) = 1\, ,
\ee
we get
\be \label{ex14}
\sum_{n\in \Zop} \tilde q^{n^2} = \sqrt 2 \,
{f_1(\tilde q) f_3(\tilde q)\over f_2(\tilde q) f_4(\tilde q)}\, .
\ee
Using Eqs. (\ref{ex11}) and (\ref{ex14}), (\ref{ex10}) then becomes 
\be \label{ex15a}
Z = 0\, .
\ee
Since the integrand of $Z$ vanishes for all $t$, this shows that there
is exact degeneracy between bosonic and fermionic open string states
at all mass level, although the brane is non-BPS. 

The critical radii where the spectrum of open strings develops exact
Bose-Fermi degeneracy correspond precisely to the values below which
the non-BPS D-brane becomes unstable against the decay into a pair of
BPS branes \cite{Sen2}. This is not a coincidence: for 
$R_i>{1\over \sqrt 2}$ the massless spectrum in light-cone gauge 
contains four bosonic states, but eight fermionic states. In order to
have Bose-Fermi degeneracy at the massless level, we need four extra
massless bosonic states; these are the would-be tachyons that
precisely become massless at the critical point.

We can use this result to conclude that when
$R_6=R_7=R_8=R_9={1\over\sqrt 2}$, the force between a pair of non-BPS
D-particles vanishes at all distances. To see this we note that if we
consider a pair of such branes separated by a distance $r$ 
in any of the non-compact directions transverse to the brane, then the
partition function of open strings stretched from one of the branes to
another is given by the same expression as (\ref{ex10}) except for an
overall extra factor of $\tilde q^{r^2/2\pi^2}$ in the integrand, 
reflecting the energy associated with the tension of the open string
stretched over a distance $r$. Thus at the critical radius the
partition function vanishes, reflecting that the potential energy
$V(r)$ between the pair of branes (which is equal to negative of the
partition function) vanishes identically for all $r$.

Since $\sum_{n_i\in \Zop} \tilde q^{2R_i^2 n_i^2}$ is a monotonically
decreasing function of $R_i$ (as $0<\tilde q<1$), we see that for 
$R_i >{1\over \sqrt 2}$ the integrand of Eq.~(\ref{ex10}) is a negative 
definite function. Thus $V(r)$ is positive definite. Furthermore since
$V(r)$ only depends on $r$ via $\tilde q^{r^2/2\pi^2}$, it follows by
the same argument that $V'(r)$ is negative, and hence that $V(r)$ is a
monotonically decreasing function of $r$. Thus for
$R_i>{1\over\sqrt 2}$, where the non-BPS brane is stable, the
interaction between a pair of such branes is repulsive at all
distances.

\section*{Acknowledgements}

I thank Oren Bergman and Ashoke Sen for many useful conversations
about issues that are covered in these lectures. I also thank Eduardo
Eyras for comments on a first version of these notes.

\noindent These lectures were given at the TMR network school on
`Quantum aspects of gauge theories, supersymmetry and quantum
gravity', Torino, 26 January  -- 2 February  2000, and at the 
`Spring workshop on Superstrings and related matters',
Trieste, 27 March -- 4 April 2000. I thank the organisers for giving
me the opportunity to present these lectures, and for organising very
successful and enjoyable meetings. I also thank the participants for
asking many useful questions that have helped to improve the
presentation of these notes.

\noindent I am grateful to the Royal Society for a University Research
Fellowship. The work was also partially supported by the PPARC SPG
programme ``String Theory and Realistic Field Theory'',
PPA/G/S/1998/00613.


\begin{thebibliography}{[20]}

\bibitem{Dai} J. Dai, R.G. Leigh, J. Polchinski, {\it New connections
between string theories}, Mod. Phys. Lett. {\bf 4}, 2073 (1989).

\bibitem{mbg1} M.B. Green, {\it Pointlike states for Type 2B
superstrings}, Phys. Lett. {\bf B329}, 435 (1994); 
{\sf hep-th/9403040}. 

\bibitem{Pol1} J. Polchinski, {\it Dirichlet branes and Ramond-Ramond
charges}, Phys. Rev. Lett. {\bf 75}, 4724 (1995); {\sf hep-th/9510017}.

\bibitem{PolCai} J. Polchinski, Y. Cai, {\it Consistency of open
superstring theories}, Nucl. Phys. {\bf B296}, 91 (1988).

\bibitem{CLNY} C.G. Callan, C. Lovelace, C.R. Nappi, S.A. Yost,
{\it Loop corrections to superstring equations of motion},
Nucl. Phys. {\bf B308}, 221 (1988).

\bibitem{Li} M. Li, {\it Boundary states of D-branes and dy-strings}, 
Nucl. Phys. {\bf B460}, 351 (1996); {\sf hep-th/9510161}.

\bibitem{GrGut} M.B. Green, M. Gutperle, {\it Light-cone
supersymmetry and $D$-branes}, Nucl. Phys. {\bf B476}, 484
(1996); {\sf hep-th/9604091}.

\bibitem{Cardy} J.L. Cardy, {\it Boundary conditions, fusion rules,
and the Verlinde formula}, Nucl. Phys. {\bf B324}, 581 (1989).

\bibitem{Lew} D. Lewellen, {\it Sewing constraints for conformal field
theories on surfaces with boundaries}, Nucl. Phys. {\bf B372}, 654
(1992). \\
J.L. Cardy, D. Lewellen, {\it Bulk and boundary operators in conformal
field theory}, Phys. Lett. {\bf B259}, 274 (1991). 

\bibitem{PSS1} G. Pradisi, A. Sagnotti, Ya. S. Stanev, {\it Planar
duality in $SU(2)$ WZW models}, Phys. Lett. {\bf B354}, 279 (1995);
{\sf hep-th/9503207}.

\bibitem{PSS2} G. Pradisi, A. Sagnotti, Ya. S. Stanev, {\it The open
descendants of non-diagonal $SU(2)$ WZW models}, Phys. Lett. {\bf B356},
230 (1995); {\sf hep-th/9506014}.

\bibitem{Love71}  C. Lovelace, {\it Pomeron form-factors and dual
Regge cuts}, Phys. Lett. {\bf B34}, 500 (1971).

\bibitem{ClaSha} L. Clavelli, J Shapiro, {\it Pomeron factorization in
general dual models}, Nucl. Phys. {\bf B57}, 490 (1973). 

\bibitem{AAGNSV} M. Ademollo, R. D'Auria, F. Gliozzi, E. Napolitano,
S. Sciuto, P. Di Vecchia, {\it Soft dilations and scale
renormalization in dual theories}, Nucl. Phys. {\bf B94}, 221 (1975). 

\bibitem{CLNY0} C.G. Callan, C. Lovelace, C.R. Nappi, S.A. Yost,
{\it Adding holes and crosscaps to the superstring}, Nucl. Phys. 
{\bf B293}, 83  (1987).  

\bibitem{BiaSag} M. Bianchi, A. Sagnotti, {\it On the systematics of
open string theories}, Phys. Lett. {\bf B247}, 517 (1990). \\ 
M. Bianchi, A. Sagnotti, {\it Twist symmetry and open string Wilson
lines}, Nucl. Phys. {\bf B361}, 519 (1991). 

\bibitem{Hor} P. Ho\v{r}ava, {\it Strings on world sheet orbifolds},
Nucl. Phys. {\bf B327}, 461 (1989).

\bibitem{BG1} O. Bergman, M.R. Gaberdiel, {\it A non-supersymmetric
open string theory and S-duality}, Nucl. Phys.~{\bf B499}, 183 (1997); 
{\sf hep-th/9701137}. 

\bibitem{BG2} O. Bergman, M.R. Gaberdiel, {\it Stable non-BPS
D-particles}, Phys. Lett.~{\bf B441}, 133 (1998);
{\sf hep-th/9806155}.

\bibitem{KT} I.R. Klebanov, A.A. Tseytlin, {\it D-branes and dual
gauge theories in type 0 Strings}, Nucl. Phys. {\bf B546}, 155
(1999); {\sf hep-th/9811035}. \\
I.R. Klebanov, A.A. Tseytlin, {\it Asymptotic freedom and infrared
behaviour in the type 0 string approach to gauge theory}, 
Nucl. Phys. {\bf B547}, 143 (1999); {\sf hep-th/9812089}. \\
I.R. Klebanov, A.A. Tseytlin, {\it A non-supersymmetric large N CFT
from Type 0 string theory}, JHEP {\bf 9903}, 015 (1999);
{\sf hep-th/9901101}.

\bibitem{Sen1} A. Sen, {\it Stable non-BPS states in string
theory}, JHEP {\bf 9806}, 007 (1998); {\sf hep-th/9803194}.

\bibitem{Sen2} A. Sen, {\it Stable non-BPS bound states of BPS
D-branes}, JHEP {\bf 9808}, 010 (1998); {\sf hep-th/9805019}.

\bibitem{Sen3} A. Sen, {\it Tachyon condensation on the brane
antibrane system}, JHEP {\bf 9808}, 012 (1998); {\sf hep-th/9805170}.

\bibitem{Sen4} A. Sen, {\it $SO(32)$ Spinors of Type I and other
solitons on brane-antibrane pair}, JHEP {\bf 9809}, 023 (1998); 
{\sf hep-th/9808141}. 

\bibitem{Sen5} A. Sen, {\it  Type I D-particle and its interactions}, 
JHEP {\bf 9810}, 021 (1998); {\sf hep-th/9809111}. 

\bibitem{Sen6} A. Sen, {\it BPS D-branes on non-supersymmetric
cycles}, JHEP {\bf 9812}, 021 (1998); {\sf hep-th/9812031}.

\bibitem{SenRev} A. Sen, {\it Non-BPS States and Branes in String
Theory}, {\sf hep-th/9904207}.

\bibitem{WittenK} E. Witten, {\it D-branes and K-theory}, JHEP 
{\bf 9812}, 019 (1998); {\sf hep-th/9810188}. 

\bibitem{Horava} P. Ho\v{r}ava, {\it Type IIA D-branes, K-theory, and
matrix theory},  Adv. Theor. Math. Phys. {\bf 2}, 1373 (1998);
{\sf hep-th/9812135}.

\bibitem{Gukov} S. Gukov, {\it K-Theory, reality, and orientifolds}, 
{\sf hep-th/9901042}.

\bibitem{BGH} O. Bergman, E.G. Gimon, P. Ho\v{r}ava, {\it Brane
transfer operations and T-duality of non-BPS states}, JHEP {\bf 9904},
010 (1999); {\sf  hep-th/9902160}.

\bibitem{stringfield} A. Sen, {\it Universality of the tachyon
potential}, JHEP {\bf 9912}, 027 (1999); {\sf hep-th/9911116}. \\
A. Sen, B. Zwiebach, {\it Tachyon condensation in string field
theory}, JHEP {\bf 0003}, 002 (2000); {\sf hep-th/9912249}. \\
N. Berkovits, A. Sen, B. Zwiebach, {\it Tachyon condensation in
superstring field theory}, {\sf hep-th/0002211}.

\bibitem{BD} J.D. Blum, K.R. Dienes, {\it Duality without
supersymmetry: the case of the $SO(16)\times SO(16)$ string}, 
Phys. Lett. {\bf B414}, 260 (1997); {\sf hep-th/9707148}.\\
J.D. Blum, K.R. Dienes, {\it Strong/weak coupling duality relations
for non-su\-per\-symmetric string theories}, Nucl. Phys. {\bf B516}, 83
(1998); {\sf hep-th/9707160}.

\bibitem{BG4} O. Bergman, M.R. Gaberdiel, {\it Dualities of Type 0
Strings}, JHEP {\bf 9907}, 022 (1999); {\sf hep-th/9906055}.

\bibitem{BK} R. Blumenhagen, A. Kumar, {\it A Note on Orientifolds and
Dualities of Type 0B String Theory}, Phys. Lett. {\bf B464}, 46 (1999);
{\sf hep-th/9906234}.

\bibitem{KKS} S.~Kachru, J.~Kumar, E.~Silverstein, {\it Vacuum Energy
Cancellation in a Non-supersymmetric String}, Phys. Rev. {\bf D59},
106004 (1999); {\sf hep-th/9807076}. 

\bibitem{KS} S.~Kachru, E.~Silverstein, {\it Self-dual
nonsupersymmetric Type II string compactifications}, JHEP {\bf 9811},
001 (1998); {\sf hep-th/9808056}. \\
S.~Kachru, E.~Silverstein, {\it On vanishing two loop cosmological
constants in nonsupersymmetric strings}, JHEP {\bf 9901}, 004 (1999);
{\sf hep-th/9810129}.

\bibitem{Harvey} J.A.~Harvey, {\it String Duality and
Non-supersymmetric Strings}, Phys. Rev. {\bf D59}, 026002 (1999); 
{\sf hep-th/9807213}.

\bibitem{Maldacena} J.M. Maldacena, {\it The large N limit of
superconformal field theories and supergravity},
Adv. Theor. Math. Phys. {\bf 2}, 231 (1998) and
Int. J. Theor. Phys. {\bf 38}, 1113 (1999); {\sf hep-th/9711200}.

\bibitem{ADS} I. Antoniadis, E. Dudas, A. Sagnotti, {\it  Brane
Supersymmetry Breaking}, Phys. Lett. {\bf B464}, 38 (1999); 
{\sf hep-th/9908023}.

\bibitem{AU} G. Aldazabal, A.M. Uranga, {\it Tachyon-free
Non-supersymmetric Type IIB Orientifolds via Brane-Antibrane Systems}, 
JHEP {\bf 9910}, 024 (1999); {\sf  hep-th/9908072}.

\bibitem{AIQ} G. Aldazabal, L.E. Ibanez, F. Quevedo, 
{\it Standard-like Models with Broken Supersymmetry from Type I String 
Vacua}, JHEP {\bf 0001}, 031 (2000); {\sf hep-th/9909172}.

\bibitem{AADDS}  C. Angelantonj, I. Antoniadis, G. D'Appollonio,
E. Dudas, A. Sagnotti, {\it Type I vacua with brane supersymmetry
breaking}, {\sf hep-th/9911081}.

\bibitem{Mukhi1} S. Mukhi, N.V. Suryanarayana, D. Tong, 
{\it Brane-Antibrane constructions}, JHEP {\bf 0003}, 015 (2000); 
{\sf hep-th/0001066}.
                                
\bibitem{Mukhi2} S. Mukhi, N.V. Suryanarayana, {\it A Stable Non-BPS
Configuration From Intersecting Branes and Antibranes},
{\sf hep-th/0003219}.

\bibitem{GabSen} M.R.~Gaberdiel, A.~Sen, {\it Non-supersymmetric
D-Brane configurations with bose-fermi de\-ge\-ne\-rate open string
spectrum}, JHEP {\bf 9911}, 008 (1999); {\sf hep-th/9908060}.

\bibitem{VecLic} P. Di Vecchia, A. Liccardo, {\it D branes in string
theory, I \& II}, {\sf hep-th/9912161} and {\sf hep-th/9912275}. 

\bibitem{BG3}  O. Bergman, M.R. Gaberdiel, {\it Non-BPS States in
Heterotic - Type IIA Duality}, {\bf JHEP 9903}, 013 (1999); 
{\sf hep-th/9901014}. 

\bibitem{FGLS} M. Frau, L. Gallot, A. Lerda, P. Strigazzi, 
{\it Stable non-BPS D-branes in type I string theory}, Nucl. Phys. 
{\bf B564}, 60 (2000); {\sf hep-th/9903123}. 

\bibitem{LR} A. Lerda, R. Russo, {\it Stable non-BPS states in string
theory: a pedagogical review}, {\sf hep-th/9905006}.

\bibitem{GabSte} M.R. Gaberdiel, B. Stefa\'nski, {\it D-branes on
orbifolds}, {\sf hep-th/9910109}, to appear in Nucl. Phys. {\bf B}.

\bibitem{DS} T. Dasgupta, B. Stefa\'nski, {\it Non-BPS States and
Heterotic - Type I' Duality},  Nucl. Phys. {\bf B572}, 95 (2000);
{\sf hep-th/9910217}. 

\bibitem{BS} T. Banks, L. Susskind, {\it Brane - Anti-Brane Forces}, 
{\sf hep-th/9511194}.

\bibitem{BCR} M. Billo', B. Craps, F. Roose, {\it On D-branes in Type
0 String Theory}, Phys. Lett. {\bf B457}, 61 (1999); 
{\sf hep-th/9902196}. 

\bibitem{MRGrev} M.R. Gaberdiel, {\it An introduction to conformal
field theory}, Rep. Prog. Phys. {\bf 63}, 607 (2000);
{\sf hep-th/9910156}.

\bibitem{Ishi} N. Ishibashi, {\it The boundary and crosscap states in
conformal field theories}, Mod. Phys. Lett. {\bf A4}, 251 (1989). 

\bibitem{OnoIshi} T. Onogi, N. Ishibashi, {\it Conformal field
theories on surfaces with boundaries and crosscaps}, Mod. Phys. Lett.
{\bf A4}, 161 (1989); erratum {\it ibid} {\bf A4}, 885 (1989).

\bibitem{HKMS} J. Harvey, S. Kachru, G. Moore, E. Silverstein, 
{\it Tension is dimension}, JHEP {\bf 0003}, 001 (2000);
{\sf hep-th/9909072}.

\bibitem{RS1} A. Recknagel, V. Schomerus, {\it D-branes in Gepner
models}, Nucl. Phys. {\bf B531}, 185 (1998); {\sf hep-th/9712186}. 

\bibitem{GS} M. Gutperle, Y. Satoh, {\it D0-branes in Gepner models
and N=2 black holes}, Nucl. Phys. {\bf B555}, 477 (1999); 
{\sf hep-th/9902120}.

\bibitem{BDLR} I. Brunner, M. Douglas, A. Lawrence, C. Romelsberger,
{\it D-branes on the quintic}, {\sf hep-th/9906200}.

\bibitem{FS} J. Fuchs, Ch. Schweigert, {\it Branes: from free fields
to general backgrounds}, Nucl. Phys. {\bf B530}, 99 (1998); 
{\sf hep-th/9712257}.

\bibitem{AS} A. Alekseev, V. Schomerus, {\it D-branes in the WZW
model}, Phys. Rev. {\bf D60}, 061901 (1999); {\sf hep-th/9812193}. 

\bibitem{ARS} A. Alekseev, A. Recknagel, V. Schomerus, 
{\it Non-commutative world-volume geometries: branes on $SU(2)$ and
fuzzy spheres}, JHEP {\bf 9909}, 023 (1999); {\sf hep-th/9908040}.

\bibitem{FFFS} G. Felder, J. Fr\"ohlich, J. Fuchs, Ch. Schweigert, 
{\it The geometry of WZW branes}, {\sf hep-th/9909030}.

\bibitem{SagCar} A. Sagnotti, {\it Open strings and their symmetry 
groups}, in: Carg\`ese '87, ``Nonperturbative quantum field theory,"
eds.: G. Mack {\it et. al.} (Pergamon Press, 1988).

\bibitem{Stra} J. Strathdee, {\it Extended Poincar\'e supersymmetry},
Int. Journ. Mod. Phys. {\bf A2}, 273 (1987).

\bibitem{Witten_bound} E. Witten, {\it Bound states of strings and
p-branes}, Nucl. Phys.~{\bf B460}, 335 (1996); 
{\sf hep-th/9510135}.

\bibitem{BG5} O. Bergman, M.R. Gaberdiel, {\it Non-BPS Dirichlet
branes}, Class. Quant. Grav. {\bf 17}, 961 (2000); 
{\sf hep-th/9908126}.

\bibitem{BG6} O. Bergman, M.R. Gaberdiel, {\it On the Consistency of
Orbifolds}, {\sf hep-th/0001130}, to appear in Phys. Lett. {\bf B}. 

\bibitem{EP} E. Eyras, S. Panda, {\it The Spacetime Life of a Non-BPS
D-particle}, {\sf hep-th/0003033}.

\bibitem{Polbook} J. Polchinski, {\it String Theory I \& II}, Cambridge
University Press (1998).

\bibitem{PolWit} J. Polchinski, E. Witten, {\it Evidence
for heterotic-type I string duality}, Nucl. Phys.~{\bf B460}, 525
(1996); {\sf hep-th/9510169}.

\bibitem{Witten1} E. Witten, {\it  String theory dynamics in various
dimensions}, Nucl. Phys.~{\bf B443}, 85 (1995); {\sf hep-th/9503124}. 

\bibitem{ks} S. Kachru, E. Silverstein, {\it On Gauge
Bosons in the Matrix Model Approach to M Theory},
Phys. Lett.~{\bf B396}, 70 (1997); {\sf hep-th/9612162}.

\bibitem{lowe} D.A. Lowe, {\it Bound states of Type I' D-particles and
enhanced gauge symmetry}, Nucl. Phys.~{\bf B501}, 134 (1997);
{\sf hep-th/9702006}.

\bibitem{bgl} O. Bergman, M.R. Gaberdiel, G. Lifschytz, {\it String
Creation and Heterotic--Type I' Duality}, Nucl. Phys.~{\bf B524},
524 (1998); {\sf hep-th/9711098}.

\bibitem{Ginsparg} P. Ginsparg, {\it Comment on toroidal
compactification of heterotic superstrings}, Phys. Rev.~{\bf D35},
648 (1987).

\bibitem{DM} M. Douglas, G. Moore, {\it D-branes, quivers, and ALE
instantons}, {\sf hep-th/9603167}.

\bibitem{Douglas} M. Douglas, {\it Enhanced gauge symmetry in M(atrix)
theory}, JHEP {\bf 9707}, 004 (1997); {\sf hep-th/9612126}. \\
D. Diaconescu, M. Douglas, J. Gomis, {\it Fractional branes and
wrapped branes}, JHEP {\bf 9802}, 013 (1998); {\sf hep-th/9712230}. \\
D. Berenstein, R. Corrado, {\it Matrix theory on ALE spaces and
wrapped membranes}, Nucl. Phys. {\bf B529}, 225 (1998); 
{\sf hep-th/9803048}.

\bibitem{Aspinwall} P.S. Aspinwall, {\it Enhanced gauge symmetries and
K3 surfaces}, Phys. Lett. {\bf B357}, 329 (1995); {\sf hep-th/9507012}.

\bibitem{dh} A.~Dabholkar, J.~Harvey, {\it Nonrenormalization of the
superstring tension}, Phys. Rev. Lett.~{\bf 63}, 478 (1989).

\bibitem{MajSen} J.~Majumder, A.~Sen, {\it Blowing up D-branes on
Non-supersymmetric Cycles}, JHEP {\bf 9909}, 004 (1999);
{\sf hep-th/9906109}.

\bibitem{EMOT53}
A.~Erd{\'e}li, W.~Magnus, F.~Oberhettinger and F.~G. Tricomi, 
{\it Higher Transcendental Functions, Vol. 2}, McGraw-Hill, (1953);
p.~354~ff.

\end{thebibliography}
\end{document}